\newlength\smallfigwidth
\newlength\figwidth
  \documentclass[aps,twocolumn,floatfix,amsmath,amssymb,showpacs,showkeys,letterpaper]{revtex4}

\newcommand{\be}{\begin{equation}}
\newcommand{\ee}{\end{equation}}
\newcommand{\ba}{\begin{align}}
\newcommand{\ea}{\end{align}}
\newcommand{\bn}{\begin{eqnarray}}
\newcommand{\en}{\end{eqnarray}}

\newcommand{\bsub}{\begin{subequations}}
\newcommand{\esub}{\end{subequations}}

\usepackage{graphicx}
\usepackage{amsmath}
\usepackage{hyperref}
\usepackage{bm}

\begin{document}

\noindent
\title{State transitions and hysteresis in a transverse magnetic island chain}
\author{Gary M. Wysin} 
\email{wysin@k-state.edu}
\homepage{http://www.phys.ksu.edu/personal/wysin}
\affiliation{Department of Physics, Kansas State University, Manhattan, KS 66506-2601}  

\date{January 31, 2025}
\vskip 0.2in
\begin{abstract}
A chain of dipole-coupled elongated magnetic islands whose long axes are oriented perpendicular to the 
chain is studied for its magnetization properties. With a magnetic field applied perpendicular to the chain,
the competition between dipolar energy, shape anisotropy, and field energy leads to three types
of uniform states with distinct magnetizations: (1) oblique to the chain, (2) perpendicular to the chain,
and (3) zero due to having alternating dipoles.  The response of these states to a slowly varying field is 
analyzed, focusing on their stability limits and related oscillation modes, and the dependencies on the 
dipolar and  anisotropy constants.  Based on identifiable transitions among the three states and their 
instability points,  the theoretically predicted zero-temperature magnetization curves show significant 
dependence on the anisotropy.  The model suggests a path for designing advanced materials with desired 
magnetic properties.  Different geometries and magnetic media for the islands are considered. 
\end{abstract}

\pacs{
75.75.+a,  
85.70.Ay,  
75.10.Hk,  
75.40.Mg   
}
\keywords{magnetics, magnetic islands, frustration, dipole interactions, metastability, magnetization, hysteresis.}
\maketitle

\section{Introduction: Magnetic island chains with three stable states.}
Arrays of magnetic islands on nonmagnetic substrates include two-dimensional artificial spin ice (ASI) 
\cite{skjaervo19,Nisoli13} and linear dipole-coupled chains \cite{Ostman18,Cisternas21}, 
that attract significant research interest associated with 
frustration effects caused by competition between shape anisotropy and dipole interactions \cite{Wang06,Nguyen17}.
In the case of square lattice ASI the ground state exhibits dipoles alternating from site to site \cite{Morgan11}
along both principal axes, but with zero net magnetization. There is also an excited remanent state
\cite{Iacocca+16,Arroo+19} achieved by removal of an applied field, which has metastable non-zero magnetization 
along a diagonal direction.  These states have been studied \cite{Iacocca+16,Arroo+19,Lasnier+20,Wysin23} for 
the small amplitude oscillations that would be present on top of their structures, which could be responsible 
for destabilizing them.  By assuming the $n$th island has a dipole moment $\vec\mu_n=\mu{\bf S}_n$ of 
fixed magnitude $\mu$ but with varying direction vector ${\bf S}_n$ (sometimes called a macrospin), 
an effective Heisenberg model \cite{Wysin+13} can be used to determine the approximate dynamics.

\begin{figure}
\includegraphics[width=\figwidth,angle=0]{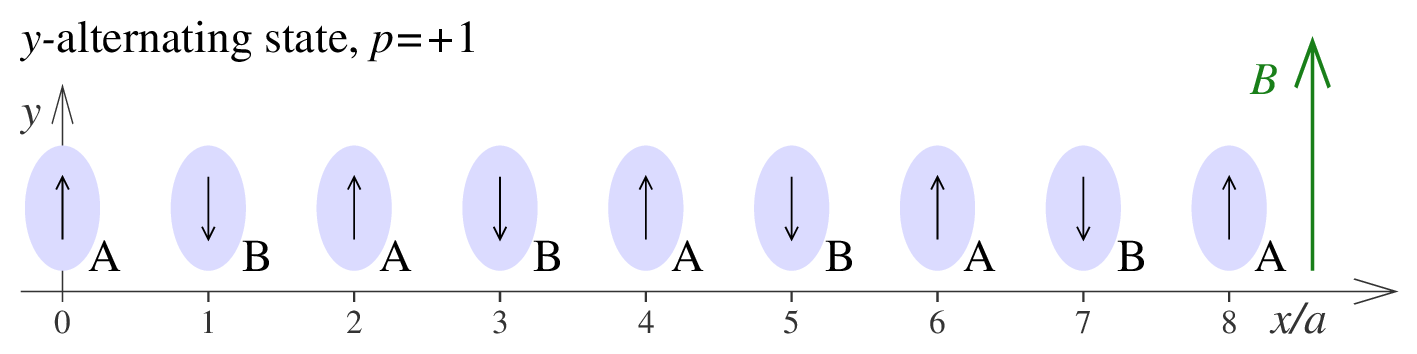}
\caption{\label{1d-islands} The linear chain of magnetic islands with their long axes perpendicular
to their separation $a$ along the chain and applied field ${\bf B}$ transverse to the chain. 
In $y$-alternating states the dipoles are uniformly aligned on each sublattice A and B.  These 
states minimize the nearest-neighbor dipolar energy ($-D$ per site) and the anisotropy energy 
($-K_1$ per site).  The other polarization $p=-1$ is obtained by reversing all dipoles.}
\end{figure}

In earlier work \cite{Wysin22} I considered a model for a dipole-coupled chain of thin islands whose long axes 
(along $\hat{y}$) are oriented perpendicular to the chain direction (along $\hat{x}$), which has some 
states reminiscent of those in square ASI.  The chain is depicted in Fig. \ref{1d-islands}. 
The islands have an elongated profile (elliptical or similar) in the $xy$-plane and a smaller thickness 
$L_z$ in the out-of-plane $z$-direction. Shape anisotropy gives an island both easy-axis anisotropy (energy 
constant $K_1$) and easy-plane anisotropy (energy constant $K_3$), that compete with long-range dipolar interactions 
and an applied field.

The main goal of this study is to determine how the possible metastable states influence the system
magnetization as a function of a magnetic field {\bf B} applied transverse to the chain. 
These phenomena could apply to chains of general island-like systems, such as patterned elements 
of Permalloy \cite{Garcia+02a} or other media, Co$_2$C nanoparticles \cite{Zhang+18} 
Fe nanoparticles \cite{Tang+15} and nanoparticle assemblies \cite{Varon+13}, and 
biomineralized magnetosomes \cite{Wittborn+99}.  With changes in island shape, spacing, and 
orientation, it may be possible to design engineered media with desired magnetic responses,  

{\em Without} an applied field, there are three types of uniform states possible \cite{Wysin22}, each
with out-of-plane components $S_n^z=0$ and two polarizations: 
(1) Two $x$-parallel states with all dipoles polarized along either $+\bm{\hat{x}}$ or $-\bm{\hat{x}}$, with minimized 
dipolar energy and high $K_1$ anisotropy energy, (2) Two $y$-parallel states with all dipoles
polarized along either $+\bm{\hat{y}}$ or $-\bm{\hat{y}}$, with minimized anisotropy energy while dipolar energy is high, 
and (3) Two $y$-alternating state, as depicted in Fig.\ \ref{1d-islands}, with dipoles along $\pm\bm{\hat{y}}$ 
but alternating at every site, where nearest-neighbor (NN) dipolar energy is low and anisotropy energy is minimized.
The $y$-alternating states are the one-dimensional analogues of the site-by-site alternating ground state
of square ASI; the $y$-parallel states are the one-dimensional analogues of remanent states of square ASI.
For brevity, I will refer to $x$-parallel as $x$-par, $y$-parallel as $y$-par and $y$-alternating as $y$-alt.

{\em With} a nonzero transverse applied field, $B\ne 0$, the dipoles of $x$-par states are tilted
by some oblique in-plane angle \cite{Wysin24} towards the field direction, pointing between the chain direction 
and the field direction.  Hence the states are renamed {\em oblique} states.  The field also strongly
affects the $y$-par states, because the one polarized parallel to the field becomes 
lower in energy than the one polarized opposite to the field.  For oblique and $y$-alt states, 
the two possible polarizations have the same energies and stability properties.

As the applied field is varied, the stability properties of these
states can change, and that will be reflected in the magnetization as a function of applied field.  
Stability was identified in Ref.\ \cite{Wysin22,Wysin24} by the absence of zero-frequency or 
imaginary-frequency normal modes of oscillation, for the whole range of allowed wave vectors.   
Stability is found to be unaffected by easy-plane anisotropy constant $K_3$, as long as it is non-negative.  
With changing $K_1$ or $B$, a particular state can change from stable to {\em metastable} or {\em unstable}.  

Here {\em unstable} means that there is at least one wave vector whose oscillation frequency is not positive; 
it could be zero, negative or even have an imaginary part.  Typically, an unstable state has a range of wave 
vectors whose mode frequencies are non-positive.  It is not possible for the state to exist for more than a 
brief instant, as it will quickly deform to some stable configuration.

I take {\em metastable} to mean that a state is locally stable against small-amplitude oscillations, with only 
positive mode frequencies, however, there exists at least one other stable state of lower energy.
By this definition, any metastable state can co-exist with other stable states, but it is in danger of
becoming unstable when the system parameters change.  To the contrary, if a change of the system parameters 
lowers its energy, that could make it become the lowest energy stable state. Parameters where a 
metastable state becomes unstable will correspond to state transitions in magnetization curves.
Thus, the analysis of the metastability properties of these states is used to characterize the magnetic responses.

The paper is organized as follows:  First, the model is further described and the energies and stability regions in 
the $(K_1,B)$  parameter space will be summarized for the three types of states.  That is followed by a discussion of
the possible allowed transitions among the states that can take place as $B$ changes.  That will be
facilitated by an effective two-sublattice potential for the system.  Based on those possible transitions,
examples of magnetization curves will be obtained and compared for different ranges of the easy-axis 
anisotropy constant $K_1$. Parameters $K_1, K_3$ and a NN-dipolar constant $D$ will be estimated for 
some different choices of island geometry and magnetic materials.

\section{Methods: The model for a linear chain of magnetic islands.}
The islands have dipoles $\vec\mu_n = \mu {\bf S}_n=\mu(S_n^x, S_n^y, S_n^z)$, where the unit direction vectors are
expressed in Cartesian or in planar spherical coordinates with an in-plane angle $\phi_n$ and an out-of-plane
angle $\theta_n$, 
\be
\label{coords}
{\bf S}_n =(\cos\theta_n \cos\phi_n, \cos\theta_n \sin\phi_n, \sin\theta_n).
\ee
As the islands are thin compared to their dimensions in the $xy$-plane, they have easy-plane anisotropy
with energy parameter $K_3 \ge 0$.  Their elongation perpendicular to the chain direction gives them
easy-axis anisotropy with energy parameter $K_1>0$.   The anisotropy energy of an island is 
taken to be
\begin{align}
H_n^K & \equiv -K_1 \left(S_n^y\right)^2 +K_3 \left(S_n^z\right)^2 
\nonumber \\
& = -K_1\cos^2\theta_n \sin^2\phi_n +K_3\sin^2\theta_n.
\end{align}
Dipoles interact with a uniform magnetic field ${\bf B}=B\mathbf{\hat{y}}$ applied transverse to the chain direction,
with energy contribution,
\be
H_n^B=-\mu B S_n^y = -\mu B \cos\theta_n\sin\phi_n.
\ee
The dipole pair interaction between dipoles at sites $n$ and $n+l$ has energy 
\begin{align}
\label{pair}
H_{n,l}^D & \equiv 
\frac{D}{l^3}
\left[ {\bf S}_n\cdot {\bf S}_{n+l} -3({\bf S}_n\cdot \hat{x})({\bf S}_{n+l}\cdot \hat{x})\right] \nonumber \\
& =  \frac{D}{l^3} \big[\sin\theta_n\sin\theta_{n+l}
+\cos\theta_n \cos\theta_{n+l} \nonumber \\ 
& \times (-2\cos\phi_n\cos\phi_{n+l}+\sin\phi_n\sin\phi_{n+l})\big].  
\end{align}
where $D$ is the nearest neighbor (NN) dipolar energy constant, defined from the permeability of
space $\mu_0$, the dipole magnitude $\mu$, and their NN separation $a$,
\be
D = \frac{\mu_0 \mu^2}{4\pi a^3}.
\ee
Then the Hamiltonian for a chain of $N$ dipoles is taken as
\be
\label{Ham}
H = \sum_{n=1}^N \left(H_n^K + H_n^B + \sum_{l=1}^R H_{n,l}^D \right).
\ee
An upper limit $R \ge 1$ on the dipole sum is the range of the dipole interactions.
$R=1$ gives the NN model, and $R\to\infty$ is referred to as the long-range-dipole (LRD) model.
For numerical results in this work, we take $R\to\infty$. The calculations are similar for finite $R$
and only result in a rescaling of frequencies and slight changes in stability limits.

The average of the summand over $n$ in expression (\ref{Ham}) is the energy per site, 
\be
U = \left\langle H_n^K + H_n^B + \sum_{l=1}^R H_{n,l}^D  \right\rangle. 
\ee
Where convenient, the energies $K_1$, $K_3$, $\mu B$, and $U$  are represented in dimensionless form
by dividing by the dipolar energy constant $D$. Those scaled energies are written using lower case symbols: 
\be
\label{parms}
k_1 \equiv \frac{K_1}{D}, \quad k_{3} \equiv \frac{K_3}{D}, \quad b \equiv \frac{\mu B}{D}, \quad u \equiv \frac{U}{D}.
\ee

\subsection{Energy and stability limits of oblique states}
In Ref.\ \cite{Wysin24} the stability limits and energies of the three types of uniform states were evaluated, 
all with global out-of-plane angle $\theta_n=0$.  The state properties are summarized here.

First, oblique states are characterized by having dipoles tilted uniformly from the $\pm \mathbf{\hat{x}}$ axis, 
with uniform spin components $S_n^y<1$ along the magnetic field direction,
\be
\label{Sny}
 S_n^y = \sin\phi_n = \frac{b/2}{3\zeta_R-k_1} \ge 0.
\ee
This depends on dipole interactions out to range $R$, involving a finite zeta-function sum,
\be
\zeta_R \equiv  \sum_{l=1}^{R} \frac{1}{l^3} .
\ee
That sum is the zero wave vector ($q=0$) limit of a truncated Clausen function \cite{Cl1,Cl2},
\be
{\rm Cl}_{R,3}(qa) \equiv \sum_{l=1}^R \frac{\cos lqa}{l^3}.
\ee
Of most interest is $R\to\infty$, in which case $\zeta_R\to \zeta(3) \approx 1.2020569$ .
The scaled per-site energy of oblique states was found to be
\be
\label{uxp}
u_{\text{oblq}} = -2\zeta_R -\frac{\left(b/2\right)^2}{3\zeta_R-k_1}.
\ee
As mentioned earlier, there are two degenerate oblique states, depending on whether the components 
$S_n^x=\cos\phi_n$ are either positive ($p=+1$) or negative ($p=-1$), as Eq.\ (\ref{Sny}) has two solutions 
for the oblique angle, $\phi_n$.

\begin{figure}[b]
\includegraphics[width=\figwidth,angle=0]{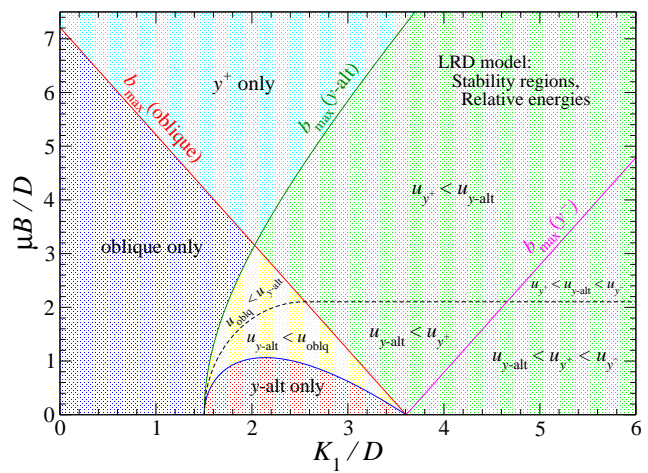}
\caption{\label{bk1} For the LRD model and any $k_3\ge 0$, the states' regions of stability are indicated
in the (uniaxial anisotropy, applied field) or $(k_1, b)$ plane, using dimensionless quantities.  
States are ranked in each region by their per-site energies from lowest to highest.  Along the dashed
black curve, $u_{y\text{-alt}}=u_{\text{oblq}}$.  Along the dashed black straight line, $u_{y\text{-alt}}=u_{y^+}$.
The plot can be mirrored across the $k_1$-axis into negative field values, while interchanging  $y^+$ and $y^-$ states. }
\end{figure}

At an upper limiting scaled field $b_{\rm max}$, $S_n^y$ reaches its maximum value, $S_n^y=1$, 
which occurs at
\be
\label{obj-Bmax}
b_{\rm max} = 2(3\zeta_R-k_1).
\ee
The parameter region for stability of oblique states is indicated in the $(k_1,b)$ plane in Fig.\ \ref{bk1}.
The maximum field $b_{\rm max}$, shown as a solid red line in Fig.\ \ref{bk1}, is where oblique states 
become identical to a $y$-par state polarized in the field direction, An oscillatory mode at $q=0$ goes to 
zero frequency at this field.  Oblique states are unstable beyond this limit.

There is also a minimum field required for stability, associated with oscillations at $qa=\pi$, given by
\be
\label{obj-Bmin}
b_{\rm min} = 2(3\zeta_R-k_1) \sqrt{\frac{k_1-\frac{5}{4}\zeta_R}{k_1+\frac{9}{4}\zeta_R}}.
\ee
The expression requires $k_1 > \frac{5}{4} \zeta_R$ and $k_1< 3\zeta_R$, or
\be
\label{Kobl}
1.50257 < k_1 < 3.606 
\ee 
for the LRD model, The minimum field in (\ref{obj-Bmin}) is represented as a solid blue curve in Fig.\ \ref{bk1}.
If $k_1$ is below $\frac{5}{4} \zeta_R$,  no minimum applied field is needed for stability.  
If the field is below $b_{\rm min}$, the calculations in Ref.\ \cite{Wysin24} showed that only $y$-alt states 
would be stable in this anisotropy range. Note that $b_{\rm min}$ and $b_{\rm max}$ do not depend on $k_3$.
Oblique states are stable predominantly for low $k_1$ anisotropy, but also in the range given in Eq.\ (\ref{Kobl})
(yellow region in Fig.\ \ref{bk1}), where the field works to stabilize oblique states.

%

\subsection{Energy and stability limits of $y$-par states}
For $y$-par states, the polarization choices $p=\pm 1$ lead to two possible uniform spin configurations,
\be
\phi_n = p \frac{\pi}{2}, \qquad S_n^y =p = \pm 1.
\ee
The per-site energies are different, because the spins are aligned with {\bf B} for one polarization but antialigned to {\bf B}
for the other polarization.  With $b=b^y$ being the $y$-component,  the per-site energy is
\be
\label{uyp}
u_{y\text{-par}} = \zeta_R-k_1-pb.
\ee
The state polarized along {\bf B} has lower energy, and better stability.  The product $pb>0$ selects
which polarization has lower energy.  Both $p$ and $b$ can be allowed to be positive or negative.  

With $b\ne 0$, the two $y$-par states are not equivalent and they need to be clearly distinguished in the discussion. 
As a result, the $p=+1$ polarization will now be referred to as the $y^{+}$ state and the $p=-1$ polarization as 
the $y^{-}$ state.  For consistency, these names are used regardless of whether $b$ is positive or negative.

Small oscillations go unstable when the energy eigenvalue $\lambda_{\phi}(q)$ associated with small in-plane 
fluctuations becomes negative at some wave vector $q$.  That eigenvalue in units of $D$ was found to be 
\be
\label{lfy}
\frac{\lambda_{\phi}(q)}{D} = -\zeta_R -2\sum_{l=1}^R \frac{\cos lqa}{l^3}+k_1+\tfrac{1}{2}pb.
\ee
Conversely, stability for these states requires positivity of $\lambda_{\phi}(q)$. One sees that $q=0$ is the
most unstable point because that gives the maximum Clausen sum and the lowest eigenvalue.  The particular 
Clausen sums \cite{Cl1,Cl2} at $qa=0$ and at $qa=\pi$, together with their LRD limits are
\begin{align}
\text{Cl}_{R,3}(0) & = \sum_{l=1}^R \frac{1}{l^3} = \zeta_R \approx 1.2020569, \nonumber \\
\text{Cl}_{R,3}(\pi) & = \sum_{l=1}^R \frac{(-1)^l}{l^3} = -\tfrac{3}{4}\zeta_R \approx -0.901542677 .
\end{align} 
Enforcing the $q=0$ requirement, $\lambda_{\phi}(0)>0$, gives an inequality that applies to both polarizations,
\be
\label{r2}
pb > 2(3\zeta_R-k_1). 
\ee
Enforcing the $qa=\pi$ requirement, $\lambda_{\phi}(qa=\pi)>0$, gives a result, 
\be
\label{r1}
pb > -\zeta_R -2 k_1.
\ee
The restriction (\ref{r2}) at $q=0$ has the greater value on the right hand side; it decides stability. 
If (\ref{r2}) is satisfied, then the relation (\ref{r1}) from $qa=\pi$ is satisfied.
There is no dependence on $k_3$.  Then for stable $y^{+}$ states the field must satisfy 
\be
\label{r2a}
b > +2(3\zeta_R-k_1), \quad y^{+} \text{ states}.
\ee
The minimum field in (\ref{r2a}) is a solid red line in the $(k_1,b)$-plane in Fig.\ \ref{bk1} (same as 
the maximum field for oblique states).  For stable $y^{-}$ states the inequality is reversed,
\be
\label{r2b}
b < -2(3\zeta_R-k_1), \quad y^{-} \text{ states}.
\ee
The maximum field in (\ref{r2b}) is a solid magenta line in Fig.\ \ref{bk1}. 
The stable regions for $y$-par states are shown in green and turquoise shading in Fig.\ \ref{bk1}.

Because the instability takes place with a $q=0$ fluctuation, the $y$-par states can destabilize only into 
oblique states or into the $y$-par state of the opposite polarization.

\subsection{Energy and stability limits of $y$-alt states}
The in-plane angles for the two polarizations of $y$-alt states are 
\be
\phi_n = (-1)^n p \frac{\pi}{2}, \qquad p=\pm 1.
\ee
Sites with even (odd) $n$ comprise the A (B) sublattice, whose spins point in opposite directions
as in Fig.\ \ref{1d-islands}.
The states have no net magnetization, but rather, $p$ refers to two choices of alternating order.  
I avoid calling it antiferromagnetic, however, because the cause is NN-dipole interactions, not 
exchange interactions. The scaled per-site energies are independent of the field, 
\be
\label{uya}
u_{y\text{-alt}} = \sum_{l=1}^R \frac{(-1)^l}{l^3} -k_1 = -\tfrac{3}{4}\zeta_R-k_1,
\ee
where the sums over even $l$ (AA and BB bonds) and odd $l$ (AB bonds) can be done separately to get this.

It is physically obvious that a strong enough field will destabilize a $y$-alt state and convert it to one 
of the  $y$-par polarizations when the field energy dominates the NN-dipolar energy,  A two-sublattice calculation 
of the small-amplitude oscillations relative to a $y$-alt state \cite{Wysin24} found that stability requires
\be
\label{b-yalt}
b < \sqrt{\left(2k_1-\tfrac{5}{2}\zeta_R\right)\left(2k_1+\tfrac{9}{2}\zeta_R\right)}.
\ee 
with no dependence on $k_3$.  The maximum field in expression (\ref{b-yalt}) is depicted in the
$(k_1,b)$-plane in Fig.\ \ref{bk1} as a solid green curve.

For all three types of planar states (all $S_n^z=0$), $k_3$ affects the 
oscillation frequencies, but not the field values where a mode acquires zero frequency.

The $y$-alt calculation showed that at the limiting field, zero-frequency oscillatory modes appear together at 
$q=0$ and $q=\pi/a$.  This suggests that the oscillations physically will connect either to the opposite $y$-alt state
(connected by a $\pi$ rotation at $q=0$ of all the spins) or to a $y$-par state (connected by a $\pi$ rotation at
$q=\pi/a$).  Below the maximum field in (\ref{b-yalt}), the $y$-alt states coexist with $y$-par
states in one region where they are stable,  and with oblique states in their stable region.
Below the minimum $b$ needed for oblique stability, Eq.\ (\ref{obj-Bmin}), $y$-alt is the only stable state.
That is, the exclusive region for $y$-alt states is 
\be
b < 2(3\zeta_R-k_1) \sqrt{\frac{k_1-\tfrac{5}{4}\zeta_R}{k_1+\tfrac{9}{4}\zeta_R}}.
\ee
This applies in the window $\tfrac{5}{4}\zeta_R < k_1 < 3\zeta_R$. 
The region in the $(k_1,b)$-plane is shown in Fig.\ \ref{bk1} with pink shading. 
A dashed black curve shows where $y$-alt and oblique states have the same per-site energy.
The dashed straight black line at $b=\frac{7}{4}\zeta_R$ shows where $y$-alt has the same per-site 
energy as the $y^{+}$ state.

\subsection{Stable and metastable regions in the anisotropy/field diagram}
Fig.\ \ref{bk1} summarizes the stability regions for each of the three types of states, in 
the (anisotropy,field)-plane for $b\ge 0$.  The diagram can be symmetrically mirrored for 
$b \le 0$ while interchanging the $y^{+}$ and $y^{-}$ labels.  

Each type of state has an exclusive region where it is the only stable state: oblique for low
anisotropy and low field (purple shading), $y^{+}$ at high positive field (turquoise shading), 
$y^{-}$ at high negative field (would be in the $b<0$ region),
and $y$-alt for $\frac{5}{4}\zeta_R < k_1 < 3\zeta_R$ and low field strength (pink shading).  

In other regions, two or more states coexist in stable form.  For those regions, the 
inequalities indicate the ranking of the per-site energies $u$ for the different states.
In the case of $y$-par and $y$-alt states, there is even a region in the lower-right area
of the plot where three possibilities coexist. A black dashed line shows where $y$-alt 
has the same per-site energy as oblique in one region and $y$-par in a second region.

In regions with more than one stable state, one has the minimum energy while the others
are metastable. If one considers changes in applied field $b$ for mapping out the magnetization 
response, the diagram helps to indicate the possible final states when one of the stability 
limit curves (in solid colors) is crossed.  That is considered next.

\section{Transitions between the states by global rotation}
In order to describe the response to a varying magnetic field and hysteresis, one needs to verify
the final state if a stability limit such as those indicated in Fig.\ \ref{bk1} is crossed.
This is obvious if the final region has only one stable state.  The difficulty arises if the final
region has two or more stable states.

\subsection{Transitions between oblique and $y$-par states}
The oblique and $y$-par states have all dipoles at the same in-plane angle $\phi_n$ from the chain axis, 
together with all out-of-plane angles $\theta_n=0$.  Their instabilities have been associated with 
small-amplitude $q=0$ fluctuations whose frequency goes to zero. Those modes are dominated by in-plane 
motions.  It is interesting to suppose that this small-amplitude deviation is extrapolated to a 
large-amplitude global rotation of the dipoles.  Setting all angles equal, $\phi_n=\phi$, the per-site 
energy obtained from the Hamiltonian becomes an effective potential for the problem, in units of $D$, 
\begin{align}
\label{uphi}
u_0(\phi) & = \sum_{l=1}^R \frac{1}{l^3} \left(-2+3\sin^2\phi\right)-k_1\sin^2\phi-b\sin\phi \nonumber \\
& = -2\zeta_R +(3\zeta_R -k_1) \sin^2\phi -b\sin\phi.
\end{align}
The $0$ subscript is used to indicate that the potential represents a $q=0$ uniform rotation of all dipoles.
The minima of this potential are the $y^{+}$ ($\phi=\pi/2$), $y^{-}$ ($\phi=-\pi/2$), or oblique (other values)
states.  The locations of the minima can be analyzed as $k_1$ and $b$ change.  If the system starts
in some minimum, and a parameter changes, the system can flow to a different minimizing state only if it
does not go over an energy barrier.  Thermal effects are not considered here.

\begin{figure}
\includegraphics[width=\figwidth,angle=0]{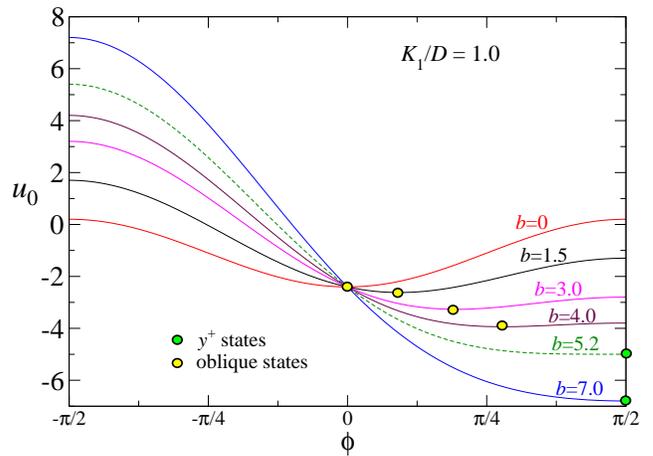}
\caption{\label{upar1} The per-site effective potential $u_0(\phi)$ in the uniform rotation assumption, Eq.\ (\ref{uphi}), 
for $k_1=1$ in the LRD model and indicated field strengths $b= \mu B/D$.  The $y^{+}$ ($\phi=\pi/2$) to 
oblique transition takes place near $b\approx 5.2$, the green-dashed curve.  Filled dots indicate energy minima, 
showing a reversible transition with changing field.  Stable $y^{-}$ states ($\phi=-\pi/2$) are not possible for 
these parameters.}
\end{figure}

A first example is shown in Fig.\ \ref{upar1}, for $k_1=1$, where the only
possible stable states are $y^{+}$ and oblique.  The filled dots show stable energy minima. 
The positions of the minima change with field $b$, indicating a transition from $y^{+}$ to
oblique states as $b$ passes below the limiting value given by Eq.\ (\ref{obj-Bmax}), $b\approx 5.2$ .
The transition is reversible: the oblique state will transform back into the $y^{+}$ state
as $b$ increases through the same limiting value.

\begin{figure}
\includegraphics[width=\figwidth,angle=0]{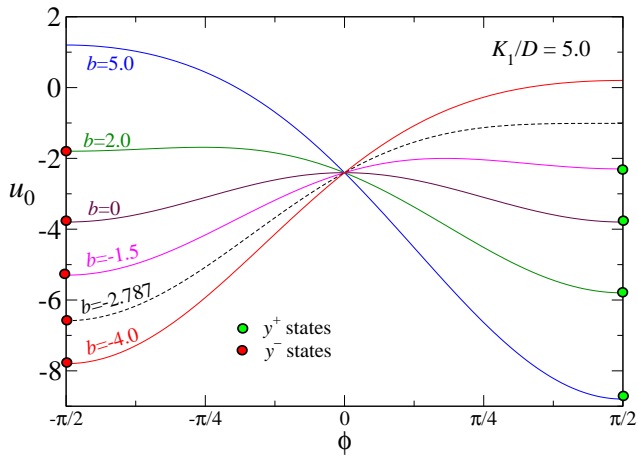}
\caption{\label{upar5} The per-site effective potential $u_0(\phi)$ in the uniform rotation assumptions, Eq.\ (\ref{uphi}), 
for $k_1=5$ in the LRD model and various indicated field strengths.  Here a mandatory non-reversible $y^{+}$ ($\phi=\pi/2$) 
to $y^{-}$ ($\phi=-\pi/2$) transition takes place near $b\approx -2.787$, where $y^{+}$ is destabilized. 
However, for $-2.787< b <+2.787$, both $y$-par states can exist.} 
\end{figure}

A second example is shown in Fig.\ \ref{upar5} for $k_1=5$. 
This anisotropy is high enough so that no oblique states are stable, and there should be only $y$-par
and $y$-alt states. The potential cannot describe $y$-alt states, but it does give an
indication of how $y^{+}$ and $y^{-}$ can coexist at some fields, while only one or the other is
stable at other fields. Note in particular at $b=-2.787$, there is no minimum near $\phi=\pi/2$, the
$y^{+}$ state, because that was its stability limit, from Eq.\ (\ref{r2a}). Indeed, if the system
starts in $y^{+}$ at $b=5.0$, and the field is reduced to below $b=-2.787$, it will transition non-reversibly 
to $y^{-}$, staying there until the field is brought back up to $b=+2.787$ (not shown), with a transition back to $y^{+}$.
The effective potential then leads to hysteresis in a magnetization curve.
 
For further analysis, one can look for extrema of $u_0(\phi)$. Its derivative with respect to $S_n^y=\sin\phi$ is
\be
\frac{du_0}{d(\sin\phi)} = 2(3\zeta_R -k_1) \sin\phi -b .
\ee
This goes to zero, locating an extremum within the range of $\phi$, when
\be
\sin\phi = \frac{b}{2(3\zeta_R-k_1)},
\ee
which is the angle for an oblique state, Eq.\ (\ref{Sny}). Starting from large $b$,
this shows a transition from $y^{+}$ into one of the polarizations of an oblique state. 
This can only happen if the formula gives $|\sin\phi| \le 1$, i.e., as $b$ is decreasing from
a large initial value.

Alternatively, if $k_1$ is large enough ($k_1>3\zeta_R$), there won't be a stable oblique state.  
Then the transition would have to be from $y^{+}$ to $y^{-}$, as seen in Fig.\ \ref{upar5}.  The final state must 
have $\phi=-\pi/2$. It is at the end of the range. That means there is no extremum within the range of $\phi$,  
and no barrier, as demonstrated in the plots of $u_0(\phi)$ for $k_1=5$, Fig.\ \ref{upar5}.  
This global rotation assumption shows the way in which the dipoles should move together to make a $y^{+}$ to $y^{-}$ 
transition at an instability limit.

\subsection{Possibility of $y$-par to $y$-alt transitions?}
In Fig.\ 16 of Ref.\ \cite{Wysin24} I showed a plot of the magnetization for this LRD model 
with $k_1=5$, as a function of the changing applied field $b$. It was drawn showing the $y^{+}$
state making a transition first to $y$-alt at $b=-2.787$, before eventually making another transition 
to $y^{-}$ near $b\approx -10$.  The assumption was that in a case where there are two possible
final states, the system would first make a transition to the one nearest in energy. That, however,
contradicts the fact that the instability in $y$-par states takes place with a $q=0$ mode,
that bears no geometric connection from the $y^{+}$ to the $y$-alt structure. Therefore,
it is important to determine by other means what the final state will be.

Stated differently, with increasing negative magnetic field, does $y^{+}$ first destabilize to 
$y$-alt and then eventually to $y^{-}$, or does $y^{+}$ destabilize directly to $y^{-}$, completely 
avoiding any $y$-alt states?

The best way to answer this might be by numerical simulation of the dynamics, starting from $y^{+}$ and setting the 
field to the destabilization limit. Alternatively, it can be answered by analysis of a two-sublattice effective
potential developed in the next section.

\section{Transitions using an effective two-sublattice potential}
To accommodate the $y$-alt states, the in-plane angles are now assumed to be uniform on each of two sublattices. 
The sites with even (odd) $n$ define the A (B) sublattice, and their in-plane angles will be $\phi_A$ ($\phi_B$). 
The effective system potential in this set of coordinates has features that indicate paths for 
transitions between pairs of oblique, $y$-par, and $y$-alt states, while keeping the planar assumption,
$\theta_n=0$ for all sites.

\subsection{Dipole interactions}
The dipolar pair interactions in Eq.\ (\ref{pair}) can be either AA bonds or AB bonds.  For an AA bond, 
$\phi_n=\phi_{n+l}=\phi_A$, where $l$ is an even nonzero number. The dipolar pair energy from (\ref{pair}) is
\be
H_{n,l,AA}^D  =  \frac{D}{l^3} (-2+3\sin^2\!\phi_A).  
\ee
The same form applies to BB bonds,
\be
H_{n,l,BB}^D  =  \frac{D}{l^3} (-2+3\sin^2\!\phi_B).  
\ee
For the inter-lattice AB or BA bonds with $l$ being odd, a pair contributes
\be
H_{n,l,AB}^D = \frac{D}{l^3} (-2\cos\phi_A\cos\phi_B+\sin\phi_A\sin\phi_B).  
\ee
I want the average per-site contribution to the energy.  Sitting at an A-site, the total dipolar energy is found by summing
over all $l\ge 1$. Summing over all even $n$ would then account for all possible AA and AB dipole pairs.
Half of the summed terms are AA and half are AB.  So an A-site's average dipolar energy contribution is
\be
U_{A}^D = \sum_{l=2,4,6...}^R H_{n,l,AA}^D +\sum_{l=1,3,5...}^R H_{n,l,AB}^D.
\ee
This involves the even/odd sums,
\begin{align}
s_{\rm e} & =\sum_{l=2,4,6...}^R \frac{1}{l^3} = \tfrac{1}{8} \zeta_R, \nonumber \\
s_{\rm o} & =\sum_{l=1,3,5...}^R \frac{1}{l^3} = \tfrac{7}{8} \zeta_R.
\end{align}
That gives the average dipole energy per  A-site,
\begin{align}
U_A^D & = \tfrac{1}{8} \zeta_R D (-2+3\sin^2\!\phi_A) \nonumber \\
& + \tfrac{7}{8} \zeta_R D (-2\cos\phi_A\cos\phi_B+\sin\phi_A\sin\phi_B).
\end{align}
The same expression applies to the dipole energy per B-site, $U_B^D$, interchanging A and B symbols.
We want the average for any site, whether A or B, which means the average of $U_A^D$ and
$U_B^D$. In units of $D$, the scaled dipole energy is $(U_A^D+U_B^D)/(2D)$ or 
\begin{align}
u_{\text{ave}}^D  = & \tfrac{1}{16}\zeta_R
(-4+3\sin^2\!\phi_A +3\sin^2\!\phi_B) \nonumber \\
& + \tfrac{7}{8} \zeta_R (-2\cos\phi_A\cos\phi_B+\sin\phi_A\sin\phi_B).
\end{align}

\subsection{The effective two-sublattice potential}
In addition to dipolar energies, there are also anisotropy and applied field terms, 
which are averaged over the sublattices.  The scaled anisotropy energy is
\be
u_{\text{ave}}^K = -\tfrac{1}{2}k_1(\sin^2\!\phi_A + \sin^2\!\phi_B)
\ee
The scaled applied field energy is
\be
u_{\text{ave}}^B = -\tfrac{1}{2}b (\sin\phi_A+\sin\phi_B).
\ee 
The total effective potential is then the sum,
\be
\label{u00}
u(\phi_A,\phi_B) = u_{\text{ave}}^D+u_{\text{ave}}^K+u_{\text{ave}}^B.
\ee
One can check that expression (\ref{u00}) reverts to the uniform rotation effective potential in Eq.\ (\ref{uphi}) 
when the sublattice angles are equal. 

Plots of the potential and its contours will be used to describe some possible transitions between states. 
It is also helpful to know its gradient function, $\vec\nabla u = (\partial u/\partial\phi_A, \partial u/\partial\phi_B)$.

The effective two-sublattice potential $u(\phi_A,\phi_B)$ gives a simplified version of the energetics,
i.e., in a limited region of the full phase space. Even so, it should contain the basic energetics needed 
to describe transformations among $y$-par, oblique, and $y$-alt states at zero temperature.  Supposing that 
the system has small damping, dynamics will take it from the current state in the direction in phase space
that has the highest downward change in energy.  Thus, it is assumed that if the system starts in a metastable
state and then a parameter such $b$ is changed, the phase point of the system will move in the direction of 
the negative gradient of the potential, $-\vec\nabla u$.

\begin{figure}
\includegraphics[width=\figwidth,angle=0]{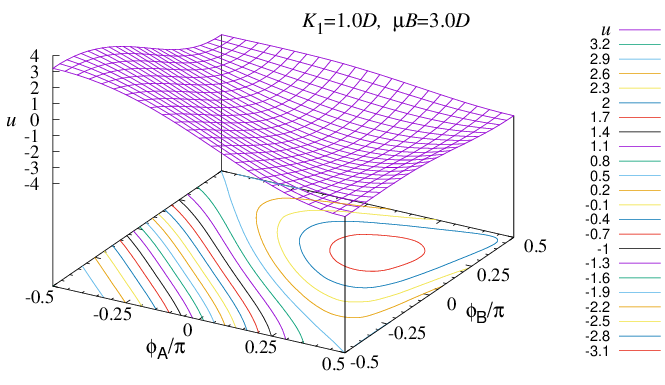}
\caption{\label{uc13} The per-site energy $u(\phi_A,\phi_B)$, in the two-sublattice LRD model, Eq.\ (\ref{u00}),
as a function of the in-plane angles $\phi_A$  and $\phi_B$ measured from the $+x$ axis, 
for $k_1=1$, $b=3$.  The contours clearly show a minimum-energy oblique state.  The $y^{+}$ state at
$\phi_A=\phi_B=\frac{1}{2}\pi$, the $y^{-}$ state at $\phi_A=\phi_B=-\frac{1}{2}\pi$, and the $y$-alt
states at the other two corners are unstable points.} \vskip 0.4cm
\end{figure}

The size of the gradient changes greatly over the relevant range of angles.  Therefore, it is
also helpful to define a {\em flow field}, 
\be
\label{flow}
\vec{f}(\phi_A,\phi_B)=-\vec\nabla u/|\vec\nabla u|, 
\ee
which shows a vector field of negative gradient arrows of fixed length.  The system point will tend to
move in the direction of $\vec{f}$.  Views of the surface $u(\phi_A,\phi_B)$, its contours, and the flow field 
give a good representation of where the system is likely to move in the limited phase space of $(\phi_A,\phi_B)$. 
These applied to three examples of the potential at different relative anisotropies $k_1$ and fields $b$,
that correspond to transitions of interest when the field changes. The examples aid in the construction of
magnetization curves vs.\ applied field.

\subsection{Low anisotropy: oblique vs. $y$-par states}
Consider first a simple case with $k_1=1$ and applied field $b=3$. These are parameters where only
an oblique state should be stable.  The energy surface $u(\phi_A,\phi_B)$ and contour plot are shown in 
Fig.\ \ref{uc13}. The stable oblique state is clearly contained inside the (red) contour near 
$\phi_A=\phi_B \approx 0.20\pi$, with $u \approx -3.1$, and it is the only minimum over the range of angles shown.  
Due to symmetry, there is no need to show angles greater than $\pi/2$ in magnitude, although the other 
oblique state would be found there.  The $y$-par and $y$-alt states are located at the corners of the range shown, 
and they are all unstable points in the potential. Plots of the gradient and flow field are not needed here
to see that the system would move towards this oblique state, if initiated elsewhere in the phase space.
If the applied field is increased to $b\approx 5.2$, the minimum of the potential moves to
$\phi_A=\phi_B=+\pi/2$, i.e., the oblique state destabilizes into the $y^{+}$ state.

%

%
\begin{figure}
\includegraphics[width=\figwidth,angle=0]{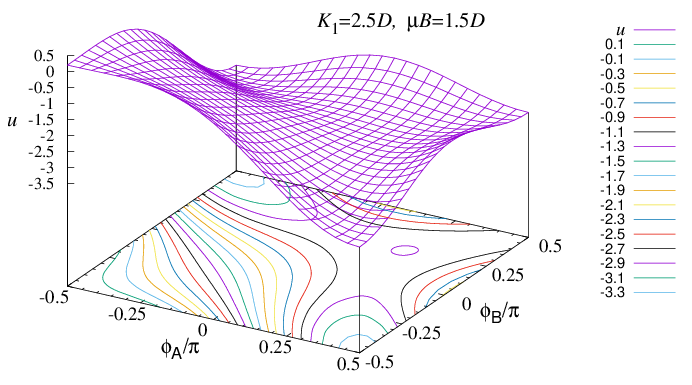}
\caption{\label{uc2515} The per-site energy $u(\phi_A,\phi_B)$, in the two-sublattice LRD model, Eq.\ (\ref{u00}),
for $k_1=2.5$, $b=1.5$ .  Note the minimum-energy oblique state at $\phi_A=\phi_B\approx 0.25 \pi$.  
The unstable $y^{+}$ state at $\phi_A=\phi_B=\frac{1}{2}\pi$ and the unstable $y^{-}$ state at 
$\phi_A=\phi_B=-\frac{1}{2}\pi$ are local energy maxima.
$y$-alt states at the other corners are stable points coexisting with the oblique state.} 
\end{figure}
\begin{figure}
\includegraphics[width=0.95\figwidth,angle=0]{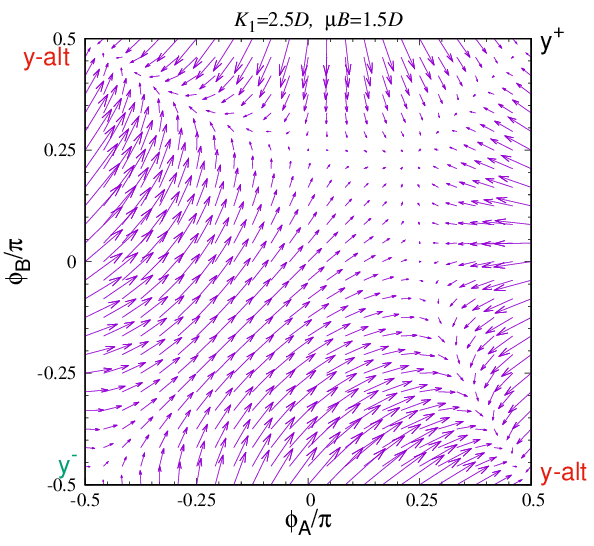}
\includegraphics[width=0.95\figwidth,angle=0]{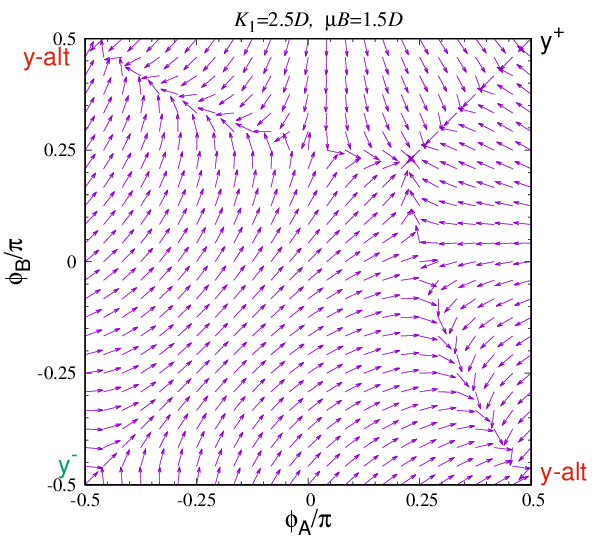}
\caption{\label{gradient2515} The negative gradient $-\vec\nabla u(\phi_A,\phi_B)$,
for $k_1=2.5$, $b=1.5$, and its unit-normalized flow pattern, Eq.\ (\ref{flow}).  The downward energy flow 
clearly moves towards a minimum-energy oblique state near $\phi_A=\phi_B\approx \pi/4$, and away from the unstable 
$y$-par states.  There is also a flow region towards stable $y$-alt states at the other corners of the range.}
\end{figure}
%
%

\subsection{Intermediate anisotropy: oblique and $y$-alt vs. $y$-par states}
The next distinct case is in a parameter region where oblique and $y$-alt coexist, while $y$-par is unstable.
Considering $k_1=2.5$, $b=1.5$, the energy surface and its contour plot are shown in Fig.\ \ref{uc2515}.
This is in the yellow region of the $(k_1,b)$-plane in Fig.\ \ref{bk1}.
There is an oblique state with a rather broad energy minimum around $\phi_A=\phi_B\approx \pi/4$. Both
$y$-par states at $\phi_A=\phi_B=\pm\pi/2$ are unstable local energy maxima.  The two $y$-alt states at
the other corners of the angular range are stable local energy minima, and slightly lower in energy than the
oblique state. This shows that the oblique state is metastable.

For comparison, the negative gradient and flow patterns are shown in Fig.\ \ref{gradient2515}.  Taken together, 
these more clearly show the flow from $y$-par states into the oblique state near $\phi_A=\phi_B\approx \pi/4$, 
where the gradient becomes weak, as indicated by short arrows there.  Less clearly, starting from other locations,
there is also flow into the $y$-alt states at $\phi_A=-\phi_B=\pm\pi/2$.  If $b$ had been slightly larger, say, $b=3$,
the $y^{+}$ state would have been stable, and the oblique state would be unstable, with $y$-alt states still stable.
Conversely, reducing $b$ towards zero will eventually eliminate the oblique state energy minimum, and the $y$-alt 
states become the only stable states, as in the pink region in Fig.\ \ref{bk1}.

\begin{figure}
\includegraphics[width=\figwidth,angle=0]{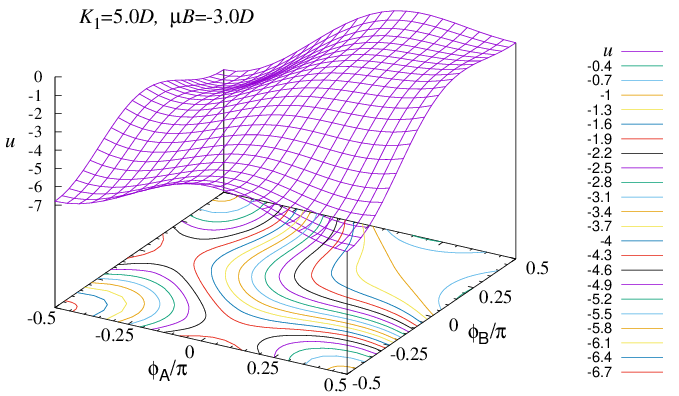}
\caption{\label{uc53} The per-site energy $u(\phi_A,\phi_B)$, in the two-sublattice LRD model, Eq.\ (\ref{u00}),
for $k_1=5.0$, $b=-3.0$, where $y^{+}$ at $\phi_A=\phi_B=\frac{1}{2}\pi$ is unstable.   
The $y^{-}$ state at $\phi_A=\phi_B=-\frac{1}{2}\pi$ is a stable energy minimum.
Both $y$-alt states at the other two corners are stable points and coexist with $y^{-}$, however, the
paths from $y^{+}$ to the $y$-alt states are highly improbable, see text and Fig.\ \ref{gradient53}.} 
\end{figure}
\begin{figure}
\includegraphics[width=\figwidth,angle=0]{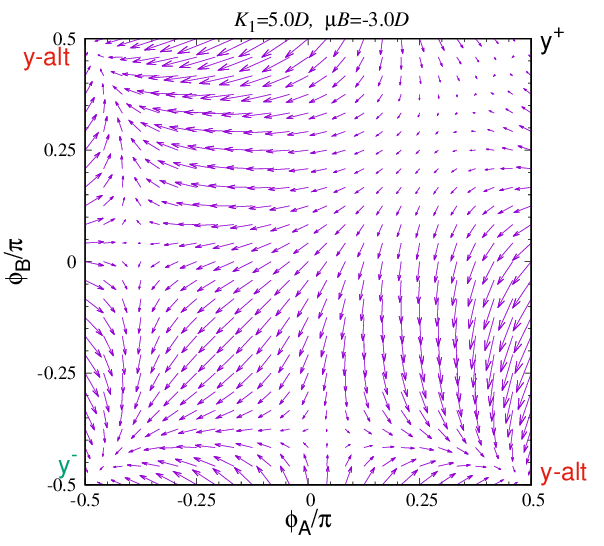}
\includegraphics[width=\figwidth,angle=0]{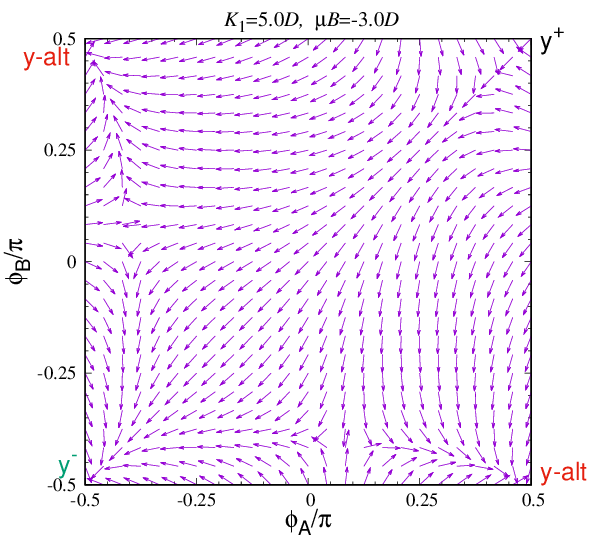}
\caption{\label{gradient53} The negative gradient $-\vec\nabla u(\phi_A,\phi_B)$ for $k_1=5.0$, $b=-3.0$, 
and its unit-normalized flow pattern, Eq.\ (\ref{flow}).  The downward energy flow from a significant 
region around the unstable $y^{+}$ state moves definitively towards the $y^{-}$ state.  The flow into the 
stable $y$-alt states does not come directly from the $y^{+}$ region.}
\end{figure}
%

\subsection{Larger anisotropy: $y$-par vs. $y$-alt states}
Consider now $k_1 > 3\zeta_R \approx 3.606$ for the LRD model.  Only $y^{+}$, $y^{-}$ and $y$-alt states 
are possible for this anisotropy, see Fig.\ \ref{bk1}.  
The $y^{+}$ state is stable for large positive fields $b \gg 1$.  If the system starts in $y^{+}$ and then the field is 
reduced or reversed, at some point $y^{+}$ will be destabilized.  That occurs at a negative (or reversed) field value, 
given by the limit in Eq.\ (\ref{r2a}).  That limiting value is indirectly represented in Fig.\ \ref{bk1} as a  
magenta line for $k_1>3.606$; that line reflected to the negative $b$ direction is the negative field limit
for a $y^{+}$ state. 

The question about the final state starting from $y^{+}$ was alluded to earlier: Upon reversing the applied field, 
as one does for a magnetization curve, will the final state be $y^{-}$ or a $y$-alt state?  

The instability of $y^{+}$ is driven by a zero-frequency mode at $q=0$, where the two sublattice angles move together,
i.e., $\phi_A=\phi_B$.  That is along a diagonal line in a plot in the $(\phi_A,\phi_B)$ plane.  The question about the
true final state can be answered from the effective potential, its contours, and especially the gradient and flow diagrams.  
To be specific, consider the case with $k_1=5.0$, for which the $y^{+}$ state destabilizes when 
$b \approx -2.787$, according to Eq.\ (\ref{r2a}), and also verified in Fig.\ \ref{upar5}.  

A plot of the potential $u(\phi_A,\phi_B)$ for field value $b=-3.0$ is adequate to see what is 
likely to happen after destabilization (it is minimally different than a plot with $b=-2.788$), see Fig.\ \ref{uc53}. 
The $y^{+}$ state is actually the highest, with energy per site $u \approx -0.798$, and clearly unstable as a local
maximum.  The $y^{-}$ state at $u \approx -6.798$ is a local energy minimum, slightly lower in energy than the 
two $y$-alt states with $u \approx -5.90$, also a local minimum. From that figure alone it looks possible
for $y^{+}$ to destabilize into either $y^{-}$ or one of the $y$-alt states.  

For another view, the gradient and flow pattern for $b=-3.0$ are plotted in Fig.\ \ref{gradient53}.  There it
becomes clear that starting from any points in a region around $y^{+}$, the flow will take the state point
quickly into the uniform rotation line, where $\phi_A=\phi_B$, and from there it goes imperatively towards the $y^{-}$ state. 
Note also the substantial saddles between the $y^{-}$ and $y$-alt states, that also help to separate the
flow pattern.  The system would have to start quite far from the $y^{+}$ state, possibly due to a large fluctuation, 
in order to end up flowing to one of the $y$-alt states.  At zero temperature, when $y^{+}$ destabilizes due to 
reversing the field, the system will transform into $y^{-}$.

Summarizing for large $k_1>3\zeta_R$, the $y^{+}$ state at a reversed field will destabilize into the 
$y^{-}$ state, not into one of the $y$-alt states.  Similarly, the $y^{-}$ state at a sufficient positive 
field will destabilize back into $y^{+}$.  The $y$-par states destabilize via a $q=0$ mode that connects
one to the other.  Based on these facts, it seems that it is difficult to move the system back into a $y$-alt 
state, at larger anisotropy, simply by changing the magnetic field, 

\section{Results: Zero-temperature magnetization curves} 
The allowed transitions between the states directly influence the average magnetic moment $\langle \mu_n^y\rangle$ 
along the field direction, as a function of that field component $B=B^y$.  For theoretical purposes, the average
transverse spin component indicates the average magnetic moment per site, and is denoted $S^y=\langle S_n^y \rangle$. 
Its saturation value is unity; multiplication by dipole magnitude $\mu$ converts it to physical units.  The variation
of $S^y$ can be predicted as the applied field $B=B^y$ is scanned through positive values and then negative values in 
a usual experiment for magnetic response.  The field is assumed to start at $B=0$, with the system in its lowest energy
state. The results depend strongly on the anisotropy strengths, $k_1$.  A vertical line at the chosen $k_1$
in the $(k_1,b)$-plane of Fig.\ \ref{bk1} passes through different possible states.  The sequence and stability 
limits of those states determine the magnetization curve. The system changes as applied field decreases may not 
reverse the changes as applied field increases, which makes hysteresis possible.

For this model, a zero-temperature non-equilibrium situation is assumed, while the applied field is changed slowly
enough so that necessary energy dissipation can occur while magnetic oscillations are damped out.  That does not mean 
the system is in the lowest energy state, but rather, it stays in any current local energy minimum state, even
metastable ones, until changing the applied field causes that state to be destabilized into some other final state.  
At a field where destabilization occurs, the system is assumed to rapidly transform into another coexisting state that is 
(1) connected geometrically to the initial state  by a zero-frequency fluctuation mode, and 
(2) the terminus of the downward energy flow pattern from the initial state.  

The geometric condition (1) is the principle that the dipoles should move, typically 
at $q=0$ or $qa=\pi$, in a manner that rotates them towards the directions they would have in the final state.
Condition (2) is even stronger; the downward energy flow must go to the actual final state, that has
to be a stable energy minimum.

For this model to make sense, some damping must be assumed, as there needs to be a process to
reduce the energy, besides any work that the applied field itself does on the system. Without damping,
strong oscillations around the final state would have to take place, carrying the released energy.  
If that kinetic-like energy is not damped out and persists, the system does not settle down into the lower 
energy local minimum, but rather maintain precession-like oscillations around it. 

Various examples at different anisotropies are discussed next, all with $k_3=0$, for the LRD model.

\begin{figure}
\includegraphics[width=\figwidth,angle=0]{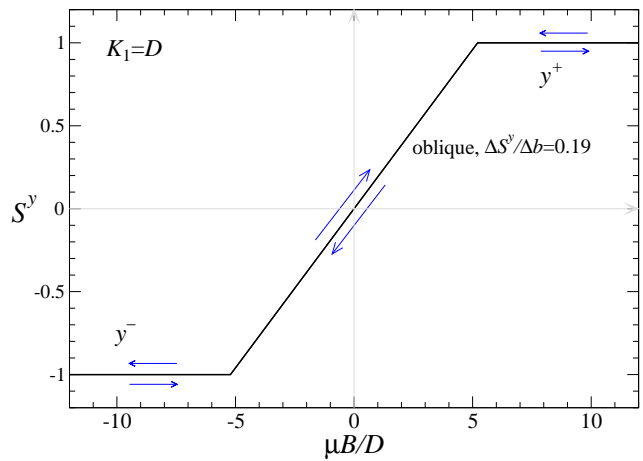}
\caption{\label{my-k1} Dimensionless magnetic moment per site, $S^y$, versus the dimensionless applied field along
$\bm{\hat{y}}$, for weak anisotropy, $k_1=1$. Oblique and $y$-par states are responsible for the shape,
as indicated.}
\end{figure}

\subsection{Magnetization variations at $k_1=1$ due to oblique--$y$-par transitions}
In the simplest example with $k_1=1$, the plot in Fig.\ \ref{bk1} shows that
only oblique and $y$-par states are possible while $b$ varies.  From Eq.\ (\ref{obj-Bmax}), the maximum 
field strength below which oblique is the only stable state is $|b|\approx 5.212$, and above which the system
in in either $y^{+}$ for $b>0$ or $y^{-}$ for $b<0$.  At that limiting field the magnetization saturates, and 
the tilt angle of the oblique state becomes $\pm \pi/2$, \textit{i.e,}, oblique becomes one of the $y$-par states. 
In the oblique state, the dimensionless magnetic moment per site  $S^y$ is given by Eq.\ (\ref{Sny}), which varies 
linearly with $b$.  
The resulting magnetization plot in Fig.\ \ref{my-k1} shows $S^y=\langle S_n^y \rangle$ versus $b$.
Once in one of the $y$-par states, the magnetization stays saturated at $S^y=\pm 1$. 
There is no hysteresis, as the system switches reversibly between oblique and $y$-par states. 
While in the oblique state, the system displays a relatively low magnetic susceptibility 
$d S^y/ d b \approx 0.19$, in dimensionless units.

\begin{figure}
\includegraphics[width=\figwidth,angle=0]{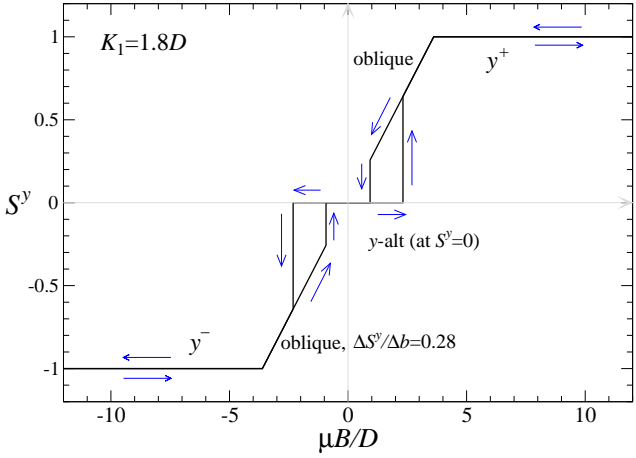}
\caption{\label{my-k1p8} Magnetization plot for mid-sized anisotropy, $k_1=1.8$, in which hysteresis is present 
due to coexistence of oblique states with $y$-alt states in some field regions.}
\end{figure}

\subsection{Magnetization variations at $k_1=1.8$}
With slightly stronger anisotropy, $k_1=1.8$, inspection of the plot in Fig.\ \ref{bk1} shows
that all three types of states are possible as $b$ is varied.  The expected zero-temperature magnetization curve is 
shown in Fig.\ \ref{my-k1p8}.  At $b=0$, only $y$-alt is stable, which is the starting point. As $b$ increases
above the lower limit of Eq.\ (\ref{obj-Bmin}), oblique is stable, however, there is no reason to transition from $y$-alt
to oblique when $b$ passes that value. Only when $b$ surpasses the $y$-alt stability limit of Eq.\ (\ref{b-yalt}) will
the $y$-alt state destabilize into an oblique  state.  With somewhat larger $b$, at the limit for stability of oblique
states, Eq.\ (\ref{obj-Bmax}), there is the transition into $y$-par and the magnetization is saturated.

Bringing the field back down, the system transitions into oblique first, which then destabilizes at its lower
field limit, Eq.\ (\ref{obj-Bmin}), into $y$-alt. These take place {\em before} the field is reversed. 
Further reduction of the field to negative values and back up reproduces the same structure as on the positive 
field side.  There are two sections with hysteresis, due to the coexistence of oblique and $y$-alt states in that 
field range. The dimensionless magnetic susceptibility of the oblique state is increased to $d S^y/ d b \approx 0.28$, 
because smaller field will saturate $S^y$, compared to that for $k_1=1$ .

\begin{figure}
\includegraphics[width=\figwidth,angle=0]{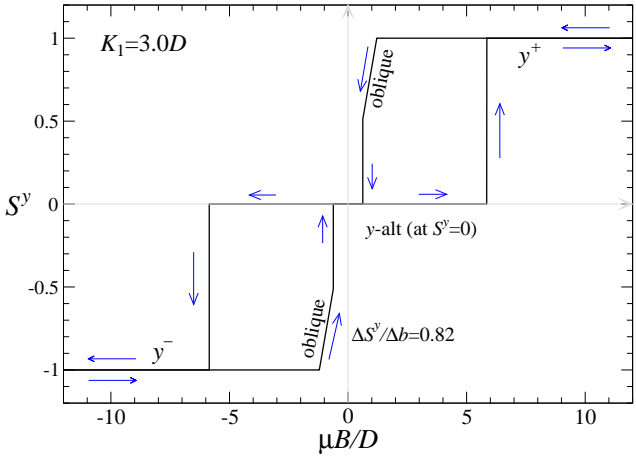}
\caption{\label{my-k3} Magnetization plot for mid-sized anisotropy, $k_1=3$, where obvious hysteresis is 
present due to coexistence of oblique or $y$-par states with $y$-alt states in some field regions.}
\end{figure}

\subsection{Magnetization variations at $k_1=3$}
A similar case to consider is raising the anisotropy to $k_1=3$, for which all three types of states 
are still possible.  The magnetization curve is shown in Fig.\ \ref{my-k3}.  Starting from $b=0$ in a
$y$-alt state, $b$ can be increased until $y$-alt destabilizes [field limit in Eq.\ (\ref{b-yalt})],
involving a direct transition into $y^{+}$.   There is no intermediate oblique state.  Reducing the
field back towards zero, the $y^{+}$ state will destabilize into oblique [field limit in Eq.\ (\ref{obj-Bmax})], 
and not go directly back to $y$-alt.  The transition to $y$-alt takes place at the lower stability limit for
oblique states, Eq.\ (\ref{obj-Bmin}), a {\em positive} field. The same transitions occur for negative field values.  
There are large hysteresis blocks where $y$-alt coexists with one of the $y$-par states, only slightly deformed from
rectangles, due to oblique being possible at lower field values.   The oblique segments are very steep, with
dimensionless magnetic susceptibility at $d S^y/ d b \approx 0.82$, precisely because only a weak magnetic field
can hold the system in a $y$-par state.

\begin{figure}
\includegraphics[width=\figwidth,angle=0]{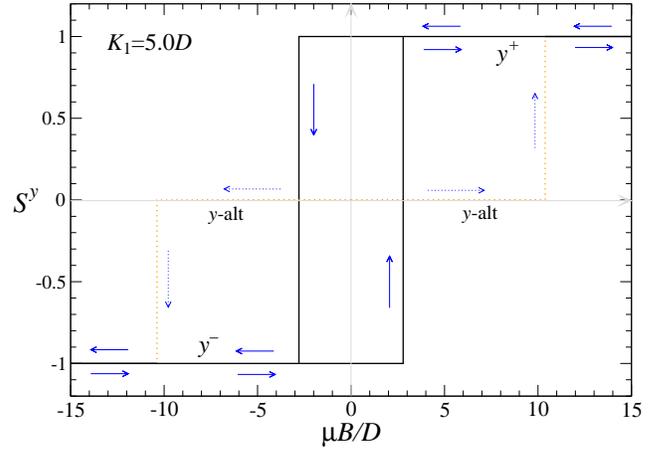} 
\caption{\label{my-k5} Magnetization plot for anisotropy $k_1=5$. Dotted lines show paths starting from a
$y$-alt state with $S^y=b=0$. Hysteresis is due to coexistence of the $y$-par states. The system can't return to
a $y$-alt state once it enters the $y$-par states, even though it has the lowest energy for low field strength.}
\end{figure}

\subsection{Magnetization variations at $k_1 > 3.606$}
The next case considered is anisotropy values $k_1 > 3.606$, for which the diagram in Fig.\ 
\ref{bk1} shows that only $y$-par and $y$-alt states are stable.  A zero-temperature magnetization curve 
for $k_1=5$ is shown in Fig.\ \ref{my-k5}; similar result holds for any $k_1>3\zeta_R$. The system is 
assumed to start in an unmagnetized minimum-energy $y$-alt state at $b=0$.  Ironically, however, applying 
an increasing and then decreasing field, the system will not return to a $y$-alt state, as long as the 
temperature is zero.   

Initially increasing $b$ on the positive side, the $y$-alt state will become unstable at the maximum field given 
in Eq.\ (\ref{b-yalt}), transforming into $y^{+}$, with positive saturated magnetization. This initial path in the 
magnetization curve is shown as a dotted curve.  Once in the $y^{+}$ state, if the field is reduced towards zero,
the system can remain in $y^{+}$ until it becomes unstable, which is a negative field value given by Eq.\ 
(\ref{obj-Bmin}).  Based on the two-sublattice effective potential $u(\phi_A,\phi_B)$ as plotted in Figs.\
\ref{uc53} and \ref{gradient53}, the system will transition from the $y^{+}$ state into the $y^{-}$ state,
with negative saturated magnetization, completely avoiding the intermediate-energy $y$-alt state, which is 
not on the downward energy flow from $y^{+}$.  

The resulting magnetization curve then contains a typical rectangular hysteresis block, where $y^{+}$ and $y^{-}$ 
states are coexisting.  The participation of $y$-alt is only possible when starting from the origin of the plot.  
At $k_1=5$, once the field surpasses a value $b\approx 10.2$,  the system is pushed into only the $y$-par states.  
It cannot return (at zero temperature) to the original $y$-alt state.  The size of the hysteresis block increases 
with increasing anisotropy parameter $k_1$.

Effectively, higher anisotropy values put the system into an exclusive two-state configuration or bistable switch.  
It switches from the initial $y$-alt state into a $y$-par state in a single-shot process, irreversibly.
One can ask, how can the system be returned to $y$-alt?  For zero temperature, and assuming $k_1$ is fixed, 
there is no way to return to $y$-alt only by changing $b$. Instead, heating the system above the Curie temperature, with 
subsequent zero field cooling, should return it to $y$-alt.    Alternatively, it may only require heating sufficiently 
to bring a $y$-par state up over one of the saddle points that links it to the $y$-alt states, as seen in 
Figs.\ \ref{uc53} and \ref{gradient53}.

%
%
\begin{figure}
\includegraphics[width=\figwidth,angle=0]{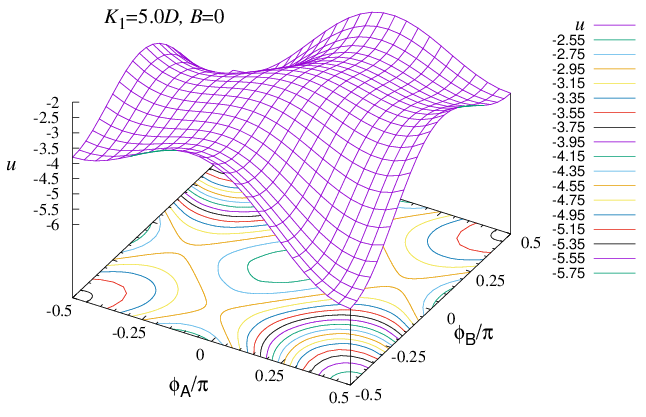}
\caption{\label{uc50} The per-site energy $u(\phi_A,\phi_B)$, in the two-sublattice LRD model, Eq.\ (\ref{u00}),
for $k_1=5.0$, $b=0.0$, where $y$-par and $y$-alt states are both stable, and connected via paths
over one of four saddle points. Compare its downward energy flow diagram in Fig.\ \ref{flow50}.} \vskip 0.4cm
\end{figure}
%

%
\begin{figure}
\includegraphics[width=\figwidth,angle=0]{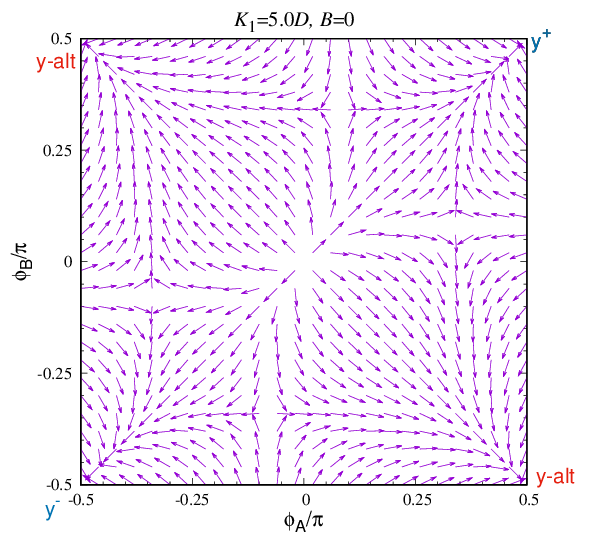}
\caption{\label{flow50} The downward energy flow pattern, Eq.\ (\ref{flow}), for $k_1=5.0$, $b=0.0$, 
where $y$-par and $y$-alt states are both stable.  The four saddle points connecting them are located at 
$(\phi_A,\phi_B) = (0.34\pi, 0.084\pi)$ and symmetry related points. See the associated surface and contour 
plots in Fig.\ \ref{uc50}.} 
\end{figure}
%

\subsection{Returning the system to $y$-alt at larger anisotropy}
For the LRD model with $k_1>3.606$, the system moves into the hysteresis loop while switching between $y^{+}$ and 
$y^{-}$ states, completely avoiding any $y$-alt state, even though $y$-alt has lower energy at $b=0$.  In the
two-sublattice effective potential $u(\phi_A,\phi_B)$, $y$-par and $y$-alt states are separated by a saddle, 
i.e., an energy barrier.  At nonzero temperature, there will be some probability of falling back into the lower-energy 
$y$-alt state, even over the barrier.  Once the temperature $T$ is high enough such that the average thermal energy 
$k_B T$ is comparable to the energy difference between $y^{+}$ and the saddle,  the $y^{+}$ state will become 
thermodynamically unstable, and a transition to $y$-alt becomes highly probable. 

Before finding the barrier, consider the energy difference between $y$-par [Eq.\ (\ref{uyp})] and $y$-alt 
[Eq.\ (\ref{uya})]  for $b=0$, as an energy for comparison. Including the unit $D$, it is
\be
\Delta U_{\text{par-alt}} = U_{y\text{-par}}-U_{y\text{-alt}} = \tfrac{7}{4}\zeta_R D.
\ee
The constant is $\frac{7}{4}\zeta(3) = 2.10$ for the LRD model.  The barrier tends to be smaller than 
$\Delta U_{\text{par-alt}}$.  

The two-sublattice potential in (\ref{u00}) is used to estimate the barrier.
For example, at $k_1=5$, Fig.\ \ref{uc50} shows a combined surface and contour plot of 
$u(\phi_A,\phi_B)$ for $b=0$.  A possible path from the $y^{+}$ minimum at $(0.5\pi,0.5\pi)$, moving upward 
in energy over a saddle at approximately $(0.34\pi, 0.084\pi)$, and then back down in energy to the 
$y$-alt state at $(0.5\pi,-0.5\pi)$ is apparent.  
There are symmetry-related paths from $y^{+}$ to the other $y$-alt state, and also the paths from $y^{-}$ to the 
$y$-alt states. The corresponding flow diagram is seen in Fig.\ \ref{flow50}.

The $y^{+}$ energy is $U_{y\text{-par}}=-3.798 D$, the saddle point energy is $U_{\text{saddle}}=-2.940 D$, and the 
$y$-alt energy is $U_{y\text{-alt}}=-5.902 D$, all energies per site. The upward barrier height is
\be
\label{saddle0}
\Delta U_{\text{saddle-par}} = U_{\text{saddle}}-U_{y\text{-par}} \approx 0.858 D.
\ee
This is about $2/5$ of the energy difference between $y$-par and $y$-alt states.  The dipole coupling $D$ is 
included because it affects the physical value for estimates for real materials, found next.

\section{Results: Anisotropy, dipole constant, and energy barrier for real islands}
I consider some real arrangements of magnetic islands made from available materials and ask:
What values of $D$, $k_1$ and $k_3$ will result, and how large is the energy barrier $\Delta U_{\text{saddle-par}}$?

These estimates are based on previous work \cite{Wysin+12} on anisotropy constants in elliptically shaped
magnetic islands. The island sizes are denoted $L_x \times L_y \times L_z$, where $L_x$ is the major diameter,
$L_y$ is the minor diameter, and $L_z$ is the vertical thickness.  Aspect ratios are defined as
\be
g_1 \equiv L_x/L_y, \qquad g_3\equiv L_x/L_z.
\ee
The anisotropy constants per unit volume, $K/V$, depend on these island aspect ratios.  
Data from Figs.\ 3 and 6 of Ref.\ \cite{Wysin+12} was used to get numerical estimates of anisotropies.

The magnetic medium for the islands is assumed to have a zero-temperature saturation magnetization $M_0$ 
and a continuum exchange stiffness $A$, which together determine an exchange length given by
\be
\lambda_{\rm ex} = \sqrt{2A/(\mu_0 M_0^2)}.   
\ee
These material parameters, together with Curie temperatures, have been summarized from Ref.\ \cite{Kuzmin+20}
and shown in Table \ref{media-table} for different soft ferromagnetic media, that could be candidates for
design of magnetic island chains.

As an elliptical cylinder, the volume of an island is 
\be
V = \pi a b L_z = \frac{\pi}{4} L_x L_y L_z .
\ee
The net dipole moment for a single aligned domain is 
\be
\mu = M_0 V.
\ee
I consider a few different island sizes, geometries, spacing $a$, and magnetic media.  
Generally, greater anisotropy constants $K_1$, $K_3$ apply for larger islands.

\subsection{Case A: 240 nm$\times$48 nm$\times$24 nm islands}  
This example geometry termed Case A is shown as purple diamonds in Fig.\ 6 of Ref.\ \cite{Wysin+12}, defined by 
$L_x=240$ nm and aspect ratios $g_1=5$, $g_3=10$.  The anisotropy constants are proportional to the volume 
and the exchange stiffness, and they are of similar size,
\begin{align}
\label{K1K3-520}
K_1 & =(0.0105/\text{nm}^{2}) A V, \nonumber \\ 
K_3 & = (0.0086/\text{nm}^{2}) A V.
\end{align}
Consider Ni$_{80}$Fe$_{20}$ Permalloy (Py) as a typical material, with saturation magnetization is $M_0 = 860$ kA m$^{-1}$, 
continuum exchange stiffness $A=13$ pJ/m, and exchange length $\lambda_{\rm ex} = 5.3$ nm.  
The island volume is 
$V = \frac{\pi}{4} (240 \text{ nm})(48 \text{ nm})(24 \text{ nm}) = 2.171 \times 10^{-22}$ m$^3$
while the dipole moment is $\mu = 1.867 \times 10^{-16}$ Am$^2$.  Then the estimated anisotropies for Py islands are
\be
K_1 =  29.6\text{ aJ}, \quad K_3 = 24.3 \text{ aJ}.
\ee
Considering separation equal to the island length, $a=240$ nm, the NN-dipole coupling is estimated as  
\be
D = \frac{\mu_0\mu^2}{4\pi a^3}= 0.252 \text{ aJ}.
\ee
Clearly it is a small energy scale compared to the anisotropies; it could be increased by using closer separation, but
at the risk of invalidating the point dipole approximation. Then in the case of Py material, the
scaled anisotropies would be
\be
k_1=K_1/D= 118, \quad k_3=K_3/D = 96.
\ee
These are somewhat large, due to the weakness of the dipole interactions relative to the anisotropies.
A real system with values of $k_1$ and $k_3$ less than 10 would be better for exhibiting the features
found in this study.
Therefore, as $A$ determines the anisotropies, and $M_0$ determines $\mu$ and hence $D$, it may be better 
to use a FM material that has lower exchange stiffness and higher saturation magnetization, if one wants 
$k_1$ and $k_3$ to be of the order of 10.

The scaled anisotropies can be expressed in terms of the exchange length, which is better for
considering different media, as follows:  
\begin{align}
\label{k1k3-eq}
k_1 & = 2\pi (0.0105) \frac{a^3}{V}\left(\frac{\lambda_{\rm ex}}{\text{1 nm}}\right)^2, \nonumber \\
k_3 & = 2\pi (0.0086) \frac{a^3}{V}\left(\frac{\lambda_{\rm ex}}{\text{1 nm}}\right)^2.
\end{align} 
These contain a numerical anisotropy factor, followed by a geometrical factor $a^3/V$, followed by a material factor of
squared exchanged length. This shows that shorter exchange length (i.e., use a different medium) and larger island 
volume will lead to weaker relative shape anisotropies.  
For this geometry, the geometrical factor is
\be 
\frac{a^3}{V} = \frac{(240 \text{ nm})^3}{2.171 \times 10^{-22} \text{ m}^3} = 63.7
\ee
One way to decrease this factor is to use the same in-plane aspect ratio ($g_1=5$) but change to thicker
islands (smaller $g_3=L_x/L_z$).   When $L_z$ is increased, the study in Ref.\ \cite{Wysin+12}
indicated that $K_1/V$ also increases while $K_3/V$ decreases.  

\begin{table}[ht]
\caption{\label{media-table} Critical temperature $T_C$, zero-temperature saturation magnetization 
$M_0$, exchange stiffness $A$, and exchange length $\lambda_{\rm ex}$ for ferromagnetic media \cite{Kuzmin+20}.}
\begin{center}
\begin{tabular}{ l c c c c }
Medium  & $T_C$ (K) &  $M_0$ (MA/m) & $A$ (pJ/m)  & $\lambda_{\rm ex}$ (nm) \\
\hline
\hline
 Py  &  872  &  0.86     & 13        &  5.3   \\
 MnB     &  567  &  0.99     & 5.9       &  3.1   \\
 Fe$_2$B &  1005  &  1.29        & 9.6       &  3.0   \\
 Fe     &  1044  &  1.75         & 21        &  3.3   \\
 Co     &  1385  &  1.45         & 56        &  6.5   \\
 Ni     &  631  &  0.51         & 15        &  9.6   \\
 EuO    &  69.6  &  1.91         & 0.77      &  0.58  \\
 Gd     &  289  &  2.12         & 2.5       &  0.94  \\
 GdZn   &  270  &  1.45         & 1.6       &  1.10  \\
 Y$_2$Fe$_{17}$ &  312  & 1.24  & 3.9 & 2.0  \\
 Y$_2$Fe$_{14}$Si$_3$ &  483  &  0.95  & 4.5  &  2.8  \\
 YFe$_{11}$Ti &  534  &  1.09   & 5.8       &  2.8   \\
\hline
\end{tabular}
\end{center}
\end{table}

\subsubsection{Using materials other than Py}
Data on other materials from Ref.\ \cite{Kuzmin+20} has been summarized in Table \ref{media-table}. If extremely
large scaled anisotropies are to be avoided, then media with exchange length smaller than that in Py may be
better suited.  For example, gadolinium (Gd) is a rare-earth element with $A=$ 2.5 pJ/m and large $M_0=$ 2.12 MA/m 
at zero temperature, and a resulting exchange length $\lambda_{\rm ex}=0.94$ nm.  
For the Case A island geometry, one gets from Eq.\ (\ref{k1k3-eq}),
\be
k_1 = 2\pi(0.0105) (63.7) (0.94)^2 = 3.7  
\ee
Now the value is barely in the range $k_1> 3\zeta(3)$.  By adjusting $a$ to a different separation, or changing
the overall size of the islands,  this can be brought into nearby desired values. But Gd has a low Curie temperature, 
$T_C = 289$ K,  so other media may be more practical for room-temperature application.

Another possible material is GdZn alloy, for which $A=1.6$ pJ/m, $M_0=1.45$ MA/m, and $\lambda_{\rm ex}=1.10$ nm.  
Formula (\ref{k1k3-eq}) gives $k_1 = 5.1$, which could be scaled up or down by changes in separation $a$ or
reducing or increasing the island volume $V$.  The Curie temperature of 270 K is still below room temperature, however. 

For the other possible media choices in Table \ref{media-table}, 
a short exchange length ($\lambda_{\rm ex}<3.0$ nm) coupled with a high Curie temperature ($T_C > 300$ K) may be best
for producing practical island arrays with moderate-sized scaled anisotropies, depending on the desired use.  The
resulting scaled anisotropies are identified in Table \ref{k1k3-table} under the Case A column.  
Some of the iron-based compounds listed have exchange lengths about 3 nm or below, and Curie temperatures above 
room temperature. That includes Fe$_2$B, YFe$_{11}$Ti, Y$_2$Fe$_{14}$Si$_3$, and Y$_2$Fe$_{17}$.  
These may be better for possible experiments and/or devices made based on this system's physics.

\subsection{Case B: 240 nm $\times$ 80 nm $\times$ 30 nm islands} 
This example geometry termed Case B is defined by $L_x=240$ nm  with aspect ratios $g_1=3$, $g_3=8$, 
with separation $a=240$ nm. The anisotropies per volume are shown as blue squares in Fig.\ 6
of Ref.\ \cite{Wysin+12}.  This has similar $K/V$ as Case A but a larger volume,
$V= 4.524\times 10^{-22} \text{ m}^3$, which will reduce the geometric factor.  
The per-volume energies in Ref.\ \cite{Wysin+12} give anisotropies
\begin{align}
K_1 &= 0.007628 AV \text{/nm}^2,  \nonumber \\
K_3 &= 0.01204 AV \text{/nm}^2 . 
\end{align}
The scaled anisotropies are found from
\begin{align}
k_1 = 2\pi(0.007528)  \frac{a^3}{V} \left(\frac{\lambda_{\rm ex}}{1 \text{ nm}}\right)^2, \nonumber \\ 
k_3 = 2\pi(0.01204)  \frac{a^3}{V} \left(\frac{\lambda_{\rm ex}}{1 \text{ nm}}\right)^2.  
\end{align}
For this geometry, the geometrical factor is
\be
\frac{a^3}{V} = \frac{(240 \text{ nm})^3}{4.524 \times 10^{-22} \text{ m}^3} = 30.6
\ee
This will reduce both scaled anisotropies.  For example,  Py islands would have $k_1=41$, while Gd
islands would have $k_1=1.29$, which is so low that only oblique and $y$-par states would be present.
The results for other media with this geometry are shown as Case B in Table \ref{k1k3-table}.  
Especially the compounds with Gd have surprisingly low scaled anisotropies.  
Y$_2$Fe$_{17}$ also has low values, and its low $T_C$ could make for interesting tests of the
changes with temperature below and above room temperature.

\subsection{Case C: 240 nm $\times$ 120 nm $\times$ 12 nm islands} 
This example geometry termed Case C is defined by $L_x=240$ nm with aspect ratios $g_1=2$, $g_3=20$, 
and separation $a=240$ nm.  Realizing that $k_3$ does not affect the stability limits of the three types 
of states, it is good to look at the geometry that has the lowest value of $K_1/V$, and ignore the 
fact that is also has one of the highest values of $K_3/V$.  The anisotropies per volume are shown as
solid black dots in Fig.\ 6 of Ref.\ \cite{Wysin+12}.  Those values give
\begin{align}
K_1 &= 0.002413 AV \text{/nm}^2,  \nonumber \\
K_3 &= 0.02535 AV \text{/nm}^2 . 
\end{align}
The volume $V=2.7143\times 10^{-22} \text{ m}^3$ is slightly greater than that for Case A.
This gives the general scaled anisotropies,
\begin{align}
k_1 = 2\pi (0.002413)  \frac{a^3}{V} \left(\frac{\lambda_{\rm ex}}{1 \text{ nm}}\right)^2, \nonumber \\ 
k_3 = 2\pi (0.02535)  \frac{a^3}{V} \left(\frac{\lambda_{\rm ex}}{1 \text{ nm}}\right)^2.  
\end{align}
The $k_1$ value is significantly smaller than in Case A and Case B.
The geometrical factor is between that of Case A and Case B,
\be
\frac{a^3}{V} = \frac{(240 \text{ nm})^3}{2.7143 \times 10^{-22} \text{ m}^3} = 50.9 
\ee
This will result in a smaller $k_1$ scaled anisotropy, but rather large $k_3$.
To compare to cases A and B, islands of Py now have $k_1=21.7$, while islands of Gd have a
very low value, $k_1=0.68$, where only oblique and $y$-par states are stable.

The results for this geometry are shown as Case C in Table \ref{k1k3-table}.
Various compounds of those listed could be used to obtain $k_1$ from just below 1 for Gd media to 8.4 for Fe. 
The values of $k_3$ are about 10.5 times larger than $k_1$, but $k_3$ does not affect the per-site
energies nor the stability limits, because all $\theta_n=0$ for the three types of states studied.
Out-of-plane motions will have a high cost in energy when $k_3$ is large, so they will be strongly
limited.

\begin{table}[h]
\caption{\label{k1k3-table} Dimensionless scaled anisotropies for 
islands made from various ferromagnetic media separated by $a=240$ nm.
Case A: 240 nm $\times$ 48 nm $\times$ 24 nm. Case B: 240 nm $\times$ 80 nm $\times$ 30 nm.
Case C: 240 nm $\times$ 120 nm $\times$ 12 nm.}
\begin{center}
\begin{tabular}{ l c c c c c c c }
\multicolumn{2}{ c}{ }  &  \multicolumn{2}{ c}{Case A} & \multicolumn{2}{ c }{Case B} & \multicolumn{2}{ c }{Case C} \\
\hline\hline
Medium  &  $\lambda_{\rm ex}$ (nm) & $k_1$  &  $k_3$ & $k_1$  &  $k_3$ & $k_1$ & $ k_3$  \\
\hline\hline
 Py  &   5.3   &  118  &   96   &  41  &  65  &  21.7  &  228  \\
 MnB     &   3.1   &  40   &  33   &  14.1 &  22.2 &  7.4  &  78  \\
 Fe$_2$B  &   3.0   & 38   &  31   & 13.2  &  21  &  6.95  &  73  \\
 Fe       &   3.3   & 46   &  37   & 15.9  &  25  &  8.41  &  88  \\
 Co      &  6.5    & 177   &  145  &   62  &  98  &  33    &  340 \\
 Ni      &  9.6    & 387   &  317  &  135  &  213 &  71    &  750 \\
 EuO     &  0.58   & 1.41  &  1.16 &  0.49 & 0.78 &  0.26  &  2.73 \\
 Gd     &   0.94   &  3.7  &  3.0  & 1.29 &  2.0  &  0.68  &  7.2  \\
 GdZn   &   1.10   & 5.1   &  4.2  & 1.77  & 2.8  &  0.93  &  9.8  \\
 Y$_2$Fe$_{17}$  & 2.0   &  16.8 &  13.8  & 5.9 & 9.2  &  3.1  &  32 \\
 Y$_2$Fe$_{14}$Si$_3$  & 2.8  & 33 & 27 &  11.5 & 18.1 &  6.1  &  64 \\
 YFe$_{11}$Ti &  2.8  & 33  &  27  & 11.5 &  18.1  &  6.1  &  64  \\
\hline
\end{tabular}
\end{center}
\end{table}

\subsection{Case C with Y$_2$Fe$_{17}$ islands}
For Case C (240 nm $\times$ 120 nm $\times$ 12 nm) islands of Y$_2$Fe$_{17}$, the scaled 
anisotropies are $k_1=3.1$ and $k_3=32$ at separation $a=240$ nm.  The magnetization curve 
does not depend on $k_3$ and should be like that in Fig.\ \ref{my-k3}, with transitions at 
slightly different scaled fields. This is a situation where the system passes through all the
states $y$-alt, $y$-par, and oblique.  In a block of the hysteresis pattern there are two sharp 
transition fields (with values for $k_1=3.1$): the first at $b_a=6.1$, where $y$-alt transitions 
into $y^{+}$,  and the second at $b_o=0.53$, where oblique transitions back down into $y$-alt.  
The physical values $B_a$ and $B_o$ of these transition fields cannot be evaluated without knowing $D$. 

With volume $V=2.7143 \times 10^{-22}$ m$^3$ and $M_0=$ 1.24 MA/m, the dipole moment is
\be
\mu = M_0 V =  3.366 \times 10^{-16} \text{ A m}^2.
\ee
The physical value of $D$ is then
\begin{align}
D & = (10^{-7}\text{ N}/\text{A}^2)(3.366 \times 10^{-16} \text{ A m}^2)^2 / (240\text{ nm})^3
\nonumber \\
& = 8.195 \times 10^{-19} \text{ J} = 0.8195 \text{ aJ}.
\end{align}
Then the transitions in the hysteresis plot are converted to physical units,
\begin{align}
B_a &= \frac{b_a D}{\mu} = \frac{6.1(0.8195 \text{ aJ})}{336.6 \text{ aA m}^2} = 14.9 \text{ mT}, \nonumber \\
B_o &= \frac{b_o D}{\mu} = \frac{0.53(0.8195 \text{ aJ})}{336.6 \text{ aA m}^2} = 1.29 \text{ mT}.
\end{align}
These fields are quite small and should be easily accessible in experiments well below $T_C=312$ K. 

\label{rescaling}
Rescaling each of the islands' dimensions and the separation all by an identical factor $h$
will rescale both $D$ and $\mu$ by $h^2$, while preserving $D/\mu$, $a^3/V$, $k_1$, $k_3$, $b_a$ and $b_o$. 
Then although the energy scale changes, the transition fields $B_a$ and $B_o$ also remain unchanged.

\subsection{Case C: $y$-par to $y$-alt energy barrier for Y$_2$Fe$_{11}$Ti}
\label{CaseC}
\begin{figure}
\includegraphics[width=\figwidth,angle=0]{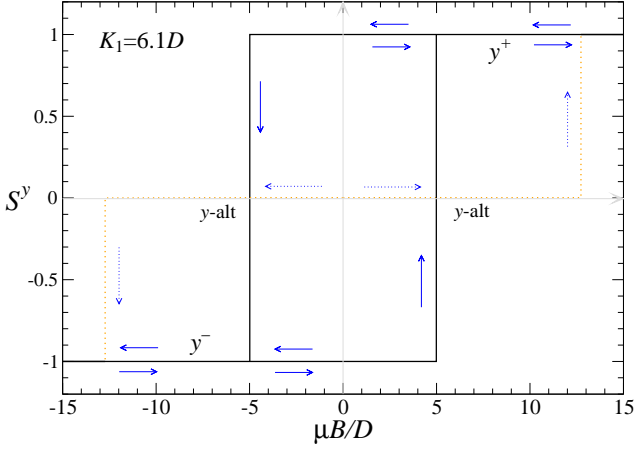}
\caption{\label{my-k6p1} Dimensionless magnetic moment per site, $S^y$, versus the dimensionless applied field along
$\bm{\hat{y}}$, for anisotropy $K_1 = 6.1D$. Dotted lines show paths starting from $S^y=B=0$.}
\end{figure}

Consider the yttrium compound with the highest $T_C$, Y$_2$Fe$_{11}$Ti, with $k_1=6.1, k_3=64$, 
for $240\times 120 \times 12$ nm$^3$ islands at separation $a=240$ nm. The $T=0$ 
magnetization plot is shown in Fig.\ \ref{my-k6p1}.  The hysteresis loop is only slightly wider 
than that for $k_1=5$ in Fig.\ \ref{my-k5}.  The transition fields in the hysteresis loop and the 
energy barrier for spontaneous transitions from $y$-par to $y$-alt are estimated here. 

The barrier and transition fields are proportional to $D$, which is first estimated.  
The volume is $V=2.7143 \times 10^{-22}$ m$^3$ and $M_0=$ 1.09 MA/m, giving the dipole moment 
\be
\mu = M_0 V =  2.96 \times 10^{-16} \text{ A m}^2.
\ee
The physical value of $D$ is
\begin{align}
D & = (10^{-7}\text{ N}/\text{A}^2)(2.96 \times 10^{-16} \text{ A m}^2)^2 / (240\text{ nm})^3 
\nonumber \\
& = 6.33 \times 10^{-19} \text{ J} = 0.633 \text{ aJ}.
\end{align}

Starting from $y$-alt, the first sharp transition takes place at $b_a=12.7$ where $y$-alt destabilizes into $y^{+}$. 
Once in the $y$-par states transitions between them take place at a scaled field magnitude $b=4.99$ from Eq.\ (\ref{r2}),
These are converted to their physical values,
\begin{align}
\label{BaBc}
B_a &= \frac{b_a D}{\mu} = \frac{12.7(0.633 \text{ aJ})}{296 \text{ aA m}^2} = 27.2 \text{ mT}. \nonumber \\
B_c &= \frac{b_c D}{\mu} = \frac{4.99(0.633 \text{ aJ})}{296 \text{ aA m}^2} = 10.7 \text{ mT}.
\end{align}
These are low fields, easily accessible in experiments. 

\begin{figure}
\includegraphics[width=\figwidth,angle=0]{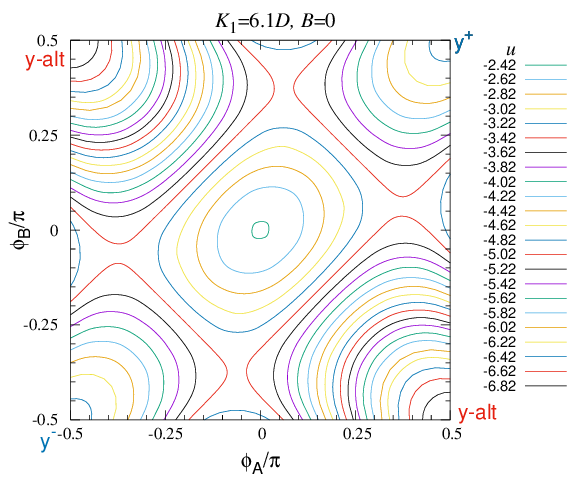}
\caption{\label{cont-k61} Effective potential contour pattern, Eq.\ (\ref{u00}), for $k_1=6.1$, $b=0.0$,
for Case C Y$_2$Fe$_{11}$Ti islands (the potential does not depend on $k_3=64$). The four saddle points connecting
$y$-par and $y$-alt states are located at $(\phi_A,\phi_B) = (0.376\pi, 0.065\pi)$ and symmetry related points.
The $y$-par, saddle, and $y$-alt per-site energies are $-4.8979 D$, $-3.4189 D$, and $-7.0015 D$, respectively.}
\end{figure}

A plot of the two-sublattice potential of Eq.\ (\ref{u00}) will indicate the energy barrier, which is 
the saddle height above the $y^{+}$ state, from which the system can relax directly down to $y$-alt.
The important points are represented in a contour plot for $u(\phi_A,\phi_B)$ for $k_1=6.1, b=0$,
in Fig.\ \ref{cont-k61}.  The $y$-par states have per-site energy $U_{y\text{-par}}=-4.8979 D$, the 
four saddle points have $U_{\text{saddle}}=-3.4189 D$, and the $y$-alt states have $U_{y\text{-alt}}=-7.0015 D$.  
Then the barrier height for a $y$-par to $y$-alt spontaneous transition is 
\be
\label{saddle-k61}
\Delta U_{\text{saddle-par}} = U_{\text{saddle}}-U_{y\text{-par}}= 1.479 D.
\ee
Combining with the result for $D$, one has
\be
\Delta U_{\text{saddle-par}} = 1.479 (0.633 \text{ aJ}) = 0.936 \text{ aJ}.
\ee
At a temperature $T=300$ K, this is about 220 times the thermal energy $k_B T$, where $k_B$
is Boltzmann's constant.  Or, the average thermal energy $k_B T$ reaches the barrier height for
$T\approx 68,000$ K.  The medium would pass above its critical temperature long before it 
goes over this barrier spontaneously. The system is very stable against any transition into $y$-alt for $T \ll T_C$. 
But note: the barrier height will be reduced to zero as the temperature reaches $T_C$. The saturation magnetization
will tend towards zero, causing $\mu\to 0$ and $D\to 0$. Heating to the Curie temperature followed 
by zero-field cooling well below $T_C$ would bring the system back to an energy-minimizing $y$-alt state.

In Section \ref{rescaling} it was noted that rescaling all the island dimensions and their
separations will shift $\mu$, $D$ and the energy barrier while preserving $k_1$ and $k_3$, 
and hence the shape of the hysteresis plot. Here I suppose the island dimensions and separation 
are all scaled down by a factor of 1/4, to 60 nm $\times$ 30 nm $\times$ 3 nm, and $a=60$ nm. 
This reduces both $\mu$ and $D$ by factors of 1/64, leading to $D=9.89$ zJ, and the barrier  becomes   
\be
\Delta U_{\text{saddle-par}} = 1.479 (9.89 \text{ zJ}) = 14.6 \text{ zJ}.
\ee
Now the temperature where thermal energy equals the barrier and spontaneous transitions can take
place becomes $T \approx 1060$ K, about twice the Curie temperature for Y$_2$Fe$_{11}$Ti.  
Still, at room temperature, over the barrier transitions are not likely. Heating the system
above $T_C$ followed by zero-field cooling will still return the system to a $y$-alt
minimum energy state. 

In terms of the hysteresis curve, the rescaling of island dimensions and separation leaves the
ratio $D/\mu$ unchanged, and the transition fields stated in (\ref{BaBc}) are unchanged.

As an alternative, modification of only the island separation or only the island dimensions 
(while preserving the aspect ratios) could be used to shift $k_1$ as well as $D$ to change
both the shape of the hysteresis curve and the overall energy scale.

\section{Discussion}
This model of a chain of elongated magnetic islands with long-range dipole interactions has three 
types of uniform states, oblique, $y$-par and $y$-alt states, all with two possible polarizations.  
Under appropriate transverse magnetic field, each state can be stable,  metastable, 
or unstable.  Metastable means the state is a local energy minimum but it has a higher 
energy than some lower stable state, to which it could decay over some energy barrier. 
At field values where a metastable state loses its stability, a transition occurs into
a lower state and manifests itself as a feature or sudden jump in the magnetization curve.  
The model allows for locating the exact conditions for these transitions to happen, depending 
only on the in-plane (easy-axis) anisotropy parameter $K_1$ and the field energy $\mu B$
measured relative to the NN dipole coupling $D$.  For various $k_1=K_1/D$, the resulting 
zero-temperature hysteresis curves have been estimated.  

For small relative anisotropy, $k_1 < \frac{5}{4}\zeta(3) \approx 1.502$, only oblique and $y$-par states are
possible, and because they are exclusive and connected at a transition field [Eq.\ (\ref{obj-Bmax})],
there is no hysteresis, as in Fig.\ \ref{my-k1}. If $k_1$ has an intermediate value,
$\frac{5}{4}\zeta(3) < k_1 < 3\zeta(3)$, the $y$-alt states dictate at weak field and complex hysteresis
patterns such as those in Figs.\ \ref{my-k1p8} and \ref{my-k3} with double blocks result.  
Finally, for large anisotropy $k_1> 3\zeta(3) \approx  3.606$, once the system has left an initial $y$-alt
state, only the two oppositely polarized $y$-par states are accessible 
(referred to as $y^{+}$ and $y^{-}$) and the hysteresis pattern has only a single block 
centered on the origin, as in Fig.\ \ref{my-k5}.  The width of that block increases with $k_1$. 

For systems with $k_1> 3\zeta(3)$ and at zero field, the $y$-alt states have the lowest energy.
Even so, once a field brings the system into $y^{+}$ or $y^{-}$, it will remain in one of those
states, even at zero field.  The energy barrier to destabilize a $y$-par state into $y$-alt 
at zero field has been estimated, and will be large compared to room-temperature thermal energy 
for examples of moderately sized islands.  The $y$-alt states will only be easily accessible afterwards 
by heating the system above the Curie temperature and subsequent zero-field cooling, which should 
return it to one of the minimum-energy $y$-alt states.

For some selection of soft FM materials and possible island sizes and spacing, estimates have been 
made for the anisotropy constants $k_1$ and $k_3$. These relative anisotropy constants are proportional to
$a^3 \lambda_{\rm ex}^2/V$, where $a$ is the island separation, $\lambda_{\rm ex}$ is the exchange length,
and $V$ is the volume of magnetic material in an island. Typically, $k_3$ tends to be larger than $k_1$,
however, only $k_1$ determines the states' stabilities and the magnetization curves.  One interesting
possibility is elliptical islands of Y$_2$Fe$_{11}$Ti of size $60 \times 30 \times 3.0$ nm$^3$
at separation $a=60$ nm, for which $k_1=6.1, k_3=64$, with transition fields at $B_a=27.2$ mT, $B_c=10.7$ mT,
and a zero-field $y$-par to $y$-alt energy barrier of 14.6 zJ ($T\approx 1060$ K).  This or other
designs could be sensitive to low fields, possibly acting as a detector or switching device.

Experimentally, it should be possible to construct magnetic island chains with properties similar to 
what is outlined here.  The behavior predicted here applies only if the islands are small enough
and well-separated, such that they have nearly uniform internal magnetization. 

\section*{Funding}
This research received no external funding.

\section*{Institutional Review Board Statement}
Not applicable.

\section*{Informed Consent Statement}
Not applicable.

\section*{Data Availability Statement}
All data generated and analyzed in this study are presented in the article.

\section*{Conflicts of Interest}
The author declares no conflicts of interest.


\end{document}